\documentclass[preprint,epsfig,graphicx,showpacs,eqsecnum,aps,amsmath,mathrsfs,eufrak,amssymb,floatfix]{revtex4}
\usepackage{graphicx}
\usepackage{epsfig}
\begin{document}

\title{Semi-microscopic description of the backbending phenomena in some deformed even-even nuclei}
\author{A. A. Raduta$^{a,b)}$ and R. Budaca $^{a)}$}

\affiliation{$^{a)}$Institute of Physics and Nuclear Engineering, RO-077125 Bucharest,
POB MG6, Romania}
\affiliation{$^{b)}$Academy of Romanian Scientists, 54 Splaiul Independentei, Bucharest 050094, Romania}
\date{\today}
\begin{abstract}
The mechanism of backbending is semi-phenomenologically investigated based on the hybridization of two rotational bands. These bands are defined by treating  a model Hamiltonian describing two interacting subsystems: a set of particles moving in a deformed mean-field and interacting among themselves through  an effective pairing force and a phenomenological deformed core whose intrinsic ground state is an axially symmetric coherent boson state.  The two components interact with each other by a quadrupole-quadrupole and a spin-spin interaction. The total Hamiltonian is considered in the space of states with good angular momentum, projected from a quadrupole deformed product function. The single-particle factor function defines the nature of the rotational bands, one corresponding to the ground band in which all particles are paired and another one built upon a $i_{13/2}$ neutron broken pair. The formalism is applied to six deformed even-even nuclei, known as being good backbenders. Agreement between theory and experiment is fairly good.
\end{abstract}
\maketitle

\renewcommand{\theequation}{1.\arabic{equation}}
\section{Introduction}
\label{sec:level1}

The anomaly in energy spacings of the rotational spectra  is still an actual subject for theoretical and experimental studies. A special attention is paid to the backbending phenomena observed in the moment of inertia dependency   on the angular velocity squared. The sudden increase for the moment of inertia at intermediate and high spin is reflected in the energy spectra by a discontinuity of the monotonous increase in energy level spacings. Backbending is a common phenomenon for many heavy and quadrupole deformed nuclei. Since its discovery \cite{371}, there were many attempts to provide a theoretical interpretation for such an anomalous behavior of nuclear energy spectra. Over the last decades studies showed that in general the backbending is a result of the crossing of the ground band ({\it g}-band) with another rotational band with a larger moment of inertia \cite{Molinari,Broglia1,Broglia2}.  The mechanism of backbending is now well known, ( i.e. $i_{13/2}$ two neutron quasiparticle alignments). The nature of the second rotational band was, however,  a question of a long standing debate, such that  a few theoretical interpretations for the origin of this band came out \cite{Ring}: (i) The second band has a deformation which is larger than that characterizing the ground band. (ii) The second band is not superfluid like the ground band but a rigid body one. (iii) The second band is built upon a broken pair with aligned individual high angular momenta. The total angular momentum of the de-paired particles being itself aligned to the core angular momentum. The later two hypotheses were the most successful ones.

The pioneering theoretical interpretation is based on the so-called Coriolis anti pairing effect (CAP) proposed by Mottelson and Valatin \cite{Mottelson}. Indeed, the Coriolis interaction violates the time reversal symmetry and consequently contributes essentially to the de-pairing process. As  a result, a phase transition from a superfluid  to an independent particle state characterized by normal fluid properties and therefore by a larger moment of inertia, takes place. 
CAP is analogous to the Meissner effect in metal superconductors \cite{Meis}. The Coriolis force is proportional to the orbital angular momentum $l$, which results in breaking first the pairs built on nucleons having largest $l$ with respect to the axis of rotation. Breaking a pair leads to a dramatic increase of the moment of inertia \cite{371} due to the large decrease of the static gap which may even vanish. Within a cranked mean-field approach, the backbending phenomenon is caused by a rearrangement of the vacuum configuration or alternatively by the crossing of the ground-state band with the lowest two quasiparticle ($2qp$) band which is often referred to as the $S$(tockholm) band \cite{371}. Information about several features like the crossing frequency, the yrast-yrare interaction and the alignment gain can be obtained from the diagram showing the $qp$ Routhians calculated at fixed deformation and for a constant pairing gap, versus the frequency \cite{218,219,373,374}. Difficulties raised due to the semiclassical nature of the cranking approach and related with the angular momentum dispersion, $\Delta({\hat J}^2)$ were discussed by several authors (see e.g., \cite{Satula}). Certainly a quantitative microscopic description of the moment of inertia in the crossing region encounters some difficulties caused by the symmetry breaking. Thus the Lipkin-Nogami method \cite{401,402} for the gauge projection was used in Ref.\cite{229} by Satula and collaborators, while the Galilean invariance \cite{408} has been restored  in Refs.\cite{410,53}. The spurious shape dependence met in treating the quadrupole-quadrupole ($QQ$) pairing interaction has been eliminated by using a double stretched $QQ$ pairing  \cite{229} and an excellent quantitative description of the moment of inertia for the so called super-deformed bands of Hg-Pb nuclei was obtained.
An alternative description of the pair breaking process \cite{Stephens} was proposed 
by Stephens and Simon, pointing to a mechanism which causes a rotational alignment of the particles from intruder orbitals. Other microscopic models reproducing qualitatively the zigzag shape of experimental plots of moment of inertia are those of Sorensen \cite{Sorensen} and Faessler \cite{Faessler, Plosz}. More recent attempts proposed approaches based on the Interacting Boson Model \cite{Iachello1,Iachello2}.

In this study we present a new and simple semi-phenomenological model for backbending which here is considered  to be the result of the band crossing mechanism between the ground band and a two quasiparticle decoupled band known as the {\it S}-band. The first backbending is known to be caused by the breaking of a neutron pair from the intruder orbital $i_{13/2}$, while the second one is due to a subsequent breaking of a proton $h_{11/2}$ pair. Our model is meant to reproduce the low spin and intermediate spin states from the yrast band, such that the scope of this paper is to describe only the first backbending. The particles from the intruder orbitals where the pair breaking occurs are treated separately from the remaining ones which define the core. The phenomenological core is described by the coherent state model \cite{Raduta1}, while the motion of the intruder particles is treated through BCS model states. A special ingredient of our formalism is that we consider here a deformed core which induces also deformed trajectories for the intruder particles. The rotational bands implied in the hybridization procedure are defined by  angular momentum projection from quadrupole deformed product states, which achieves the coupling of the single-particle degrees of freedom to a quadrupole deformed core and provides states with good angular momentum. The basis states are deformed and non-orthogonal. 
 From this basis an orthogonal basis is obtained which is further used to diagonalize the model Hamiltonian.
 The yrast energies are defined by the lowest eigenvalues of the model Hamiltonian in the orthogonal basis. This formalism was applied to six even-even nuclei from the rare earth region with $N=90-94$, which exhibit backbending behavior in their moment of inertia plots. The numerical applications reproduce quite well the experimental data and also provide some useful information regarding the rotational alignment of particles moving in the intruder orbitals.

Results are presented according to the following plan.
In the next section we describe the projected nonorthogonal basis states from which emerges the orthogonal basis. The model Hamiltonian is introduced in Sec.III where  some of its single-particle and collective features are discussed. The matrix elements of the model Hamiltonian in the non-orthogonal basis are analytically given. Section IV is devoted to deriving the final orthogonal basis and to the band hybridization procedure. Numerical results are analyzed in  Sec. V. Conclusions  and the future perspectives of the model are  discussed
in Section VI.

\renewcommand{\theequation}{2.\arabic{equation}}
\section{Description of the backbending phenomena in a restricted model space}
\label{sec: level2}

As we have already mentioned a particle-core interacting Hamiltonian will be treated in a space of spherical projected particle-core states. We start by giving the necessary details for the projection procedure and therefore about the model states. The collective factor state is a coherent state for the quadrupole bosons $b_{20}^{\dagger}$, while the single-particle component is a deformed BCS state describing a set of paired nucleons which are moving in a deformed mean-field \cite{Bes}:
\begin{equation}
\Psi\equiv\psi_{f}\psi_{b}=|BCS\rangle_{d}e^{d(b_{20}^{\dagger}-b_{20})}|0\rangle_{b}.
\end{equation}
In the above equation, $|0\rangle_{b}$ stands for the boson vacuum state, and $d$ is a real parameter which simulates the nuclear deformation. The low index of the BCS state suggests that this  is deformed. The projected states are obtained, in the
usual manner, by acting on the state (2.1) with the Hill-Wheeler projection operator
\begin{equation}
P_{MK}^{J}=\frac{2J+1}{8\pi^{2}}\int D_{MK}^{J*}\hat{R}(\Omega)d\Omega.
\end{equation}
In what follows, we shall present the angular momentum projection procedure for each factor state from (2.1).

\subsection{Angular momentum projection of the coherent state}
The projection of the quadrupole coherent state was presented by one of the authors (A. A. Raduta), in collaboration, in Ref.\cite{Raduta0}. The main results are as follows: 
\begin{equation}
\phi_{J}^{(g)}=\mathcal{N}_{J}^{(g)}P_{M0}^{J}\psi_{b}.
\end{equation}
The matrix elements of any boson Hamiltonian between these projected states can be analytically expressed in terms of the norms:
\begin{equation}
(\mathcal{N}_{J}^{(g)})^{-2}=(2J+1)e^{-d^{2}}I_{J}^{(0)},
\end{equation}
where $I_{J}^{(k)}$ stands for  the overlap integrals
\begin{equation}
I_{J}^{(k)}=\int_{0}^{1}P_{J}(y)\left[P_{2}(y)\right]^{k}e^{xP_{2}(y)}dy,\,\,\textrm{with}\,\,x=d^{2},
\end{equation}
and $P_{J}(y)$ denotes the  Legendre polynomial of the rank $J$. These integrals have been analytically calculated in Ref.\cite{Raduta0,Raduta1}.

\subsection{Spherical projected states from a deformed BCS state}  

The projection procedure adopted in this paper for  the deformed BCS state is that formulated by 
Kelemen and Dreizler in Ref.\cite{Kelemen}. Thus, the fermionic function $\psi_{f}$ is expressed
 first  as a linear combination of states with a definite angular momentum
\begin{equation}
\psi_{f}\equiv|BCS\rangle_{d}=\sum_{J}C_{J}|J,0\rangle,
\end{equation}
and then the projection operator selects only the component of the desired angular momentum:
\begin{equation}
\Psi_{JM}^{(f)}=\mathcal{N}_{J}^{(f)}P_{M0}^{J}\psi_{f}=\mathcal{N}_{J}^{(f)}C_{J}|J,M\rangle.
\end{equation}
The projected function is normalized to unity, and consequently we have:
\begin{equation}
\mathcal{N}_{J}^{(f)}=C_{J}^{-1}.
\end{equation}
In this way the amplitude $C_{J}$ can be expressed as
\begin{equation}
|C_{J}|^{2}=\langle\psi_{f}|P^{J}_{00}|\psi_{f}\rangle=\frac{2J+1}{8\pi^{2}}\int_{0}^{\pi}P_{J}(\cos{\beta})\langle\psi_{f}|e^{-i\beta\hat{J}_{y}}|\psi_{f}\rangle\sin\beta d\beta.
\end{equation}
The calculation of $C_{J}$ is very much simplified if the projection operator 
$P^{J}_{00}$ is expressed as a finite sum of particular rotation operators \cite{Kelemen}:
\begin{equation}
P_{00}^{J}=\frac{2J+1}{M+1}\left[A_{1}(M,J)+2\sum_{n=1}^{M/2}\tilde{B}_{n}(M,J)e^{-i\frac{\pi}{M+1}n\hat{J}_{y}}\right],
\end{equation}
where $J$ is even and $M=J_{max}$. On the other hand the terms $A_{1}(M,J)$ and 
$\tilde{B}_{n}(M,J)$ can be also analytically calculated. Their expressions are given in Appendix A.

The calculation of the amplitudes $C_{J}$  is thus reduced  to finding the matrix element 
$\langle\psi_{f}|e^{-i\frac{\pi}{M+1}n\hat{J}_{y}}|\psi_{f}\rangle$. For this purpose, the fermionic wave function is written as a sum of components with determined the number of particle pairs:
\begin{eqnarray}
|BCS\rangle_{d}&=&\prod_{m>0}\left(U_{jm}+V_{jm}c_{jm}^{\dagger}c_{j-m}^{\dagger}(-)^{j-m}\right)|0\rangle\nonumber\\
&=&\left(\prod_{m>0}U_{jm}\right)\left(|0\rangle+\sum_{k_{1}>0}\frac{V_{jk_{1}}}{U_{jk_{1}}}c_{jk_{1}}^{\dagger}c_{j-k_{1}}^{\dagger}(-)^{j-k_{1}}\right.\nonumber\\
&&\left.+\sum_{k_{1}<k_{2}}\frac{V_{jk_{1}}V_{jk_{2}}}{U_{jk_{1}}U_{jk_{2}}}c_{jk_{1}}^{\dagger}c_{j-k_{1}}^{\dagger}(-)^{j-k_{1}}c_{jk_{2}}^{\dagger}c_{j-k_{2}}^{\dagger}(-)^{j-k_{2}}+\ldots\right)\nonumber\\
&\equiv&\sum_{N_{p}}C_{N_{p}}|N_{p}\rangle,
\end{eqnarray}
where $|N_{p}\rangle$ are states with $N_{p}$ pairs of particles and $\{U,V\}$ are defining the Bogoliubov-Valatin (BV) transformation from the particle to the quasiparticle representation.
  It can be shown that the matrix element of the rotation operator $e^{-i\frac{\pi}{M+1}n\hat{J}_{y}}$ on states $|N_{p}\rangle$ can be expressed in the form of a determinant of rank $2N_{p}$ of reduced Wigner functions $d_{k_{1},k_{2}}^{j}(\beta_{N_{p}}^{n})$ with  the argument:
\begin{equation}
\beta_{N_{p}}^{n}=\frac{\pi\cdot n}{M+1},
\label{beta}
\end{equation}
where $M$ is the maximum angular momentum realized by the set of $N_{p}$ pairs of particles. For illustration we will present here only the case for two pairs:
\begin{eqnarray}
&\langle0|(-)^{j-k_{2}}c_{j-k_{2}}c_{jk_{2}}(-)^{j-k_{1}}c_{j-k_{1}}c_{jk_{1}}e^{-i\beta_{2}^{n}\hat{J}_{y}}c_{jk'_{1}}^{\dagger}c_{j-k'_{1}}^{\dagger}(-)^{j-k'_{1}}c_{jk'_{2}}^{\dagger}c_{j-k'_{2}}^{\dagger}(-)^{j-k'_{2}}|0\rangle=\nonumber\\
&=(-)^{(k_{1}+k_{2}+k'_{1}+k'_{2})}\det{\left(\begin{array}{cccc}d_{k_{1},k'_{1}}^{j}(\beta_{2}^{n})&d_{k_{1},-k'_{1}}^{j}(\beta_{2}^{n})&d_{k_{1},k'_{2}}^{j}(\beta_{2}^{n})&d_{k_{1},-k'_{2}}^{j}(\beta_{2}^{n})\\
d_{-k_{1},k'_{1}}^{j}(\beta_{2}^{n})&d_{-k_{1},-k'_{1}}^{j}(\beta_{2}^{n})&d_{-k_{1},k'_{2}}^{j}(\beta_{2}^{n})&d_{-k_{1},-k'_{2}}^{j}(\beta_{2}^{n})\\
d_{k_{2},k'_{1}}^{j}(\beta_{2}^{n})&d_{k_{2},-k'_{1}}^{j}(\beta_{2}^{n})&d_{k_{2},k'_{2}}^{j}(\beta_{2}^{n})&d_{k_{2},-k'_{2}}^{j}(\beta_{2}^{n})\\
d_{-k_{2},k'_{1}}^{j}(\beta_{2}^{n})&d_{-k_{2},-k'_{1}}^{j}(\beta_{2}^{n})&d_{-k_{2},k'_{2}}^{j}(\beta_{2}^{n})&d_{-k_{2},-k'_{2}}^{j}(\beta_{2}^{n})\end{array}\right)}.
\label{matwig}
\end{eqnarray}
The largest angular momentum $M$ in the configuration $\left(j\right)^{2N_{p}}$, where $j$ is the angular momentum of the individual particles, is not necessary $2N_{p}j$ because of the Pauli principle constraint. Group theory provides a simple formula for the upper limit of the total angular momentum of a given configuration, which moreover takes care of the Pauli principle \cite{Hamermesh}:
\begin{gather}
M=N_{p}(2j-2N_{p}+1).
\end{gather}
In Table I the largest angular momenta achieved for all possible numbers of neutron pairs in the intruder orbital $i_{13/2}$, are listed. 
\begin{table}[h!]
\caption{Maximum angular momentum $M$ achieved for a given number of particles, each carrying an angular momentum of $j=13/2$.}
\begin{tabular}{|c|cccccccc|}
\hline
Number of particles, $2N_{p}$&~~~0~~~&~~~2~~~&~~~4~~~&~~~6~~~&~~~8~~~&~~10~~~&~~12~~~&~~14~~~\\
\hline
Maximum angular momentum $M$&0&12&20&24&24&20&12&0\\
\hline
\end{tabular}
\end{table}
The results from this table are used to calculate the arguments $\beta^n_{N_p}$ by means of 
Eq.(\ref{beta}) and then the overlaps of the type (\ref{matwig}). With all these done the average of the rotation operator $e^{-i\frac{\pi}{M+1}n\hat{J}_{y}}$ with $|BCS\rangle_{d}$ is readily obtained.
Finally the projected particle-core function is written in the form:
\begin{equation}
\Psi_{JM}^{(1)}=\mathcal{N}_{J}^{(1)}P_{M0}^{J}|BCS\rangle_{d}\psi_{g}=\mathcal{N}_{J}^{(1)}\sum_{J_{f}J_{c}}\frac{C^{J_{f}J_{c}J}_{0\,\,0\,\,0}}{N_{J_{f}}^{BCS}N_{J_{c}}^{(g)}}\left[\psi_{J_{f}}^{BCS}\phi_{J_{c}}^{(g)}\right]_{JM},
\label{Psi1}
\end{equation}
with the normalization factor
\begin{equation}
\left(\mathcal{N}_{J}^{(1)}\right)^{-2}=\sum_{J_{f}J_{c}}\left(\frac{C^{J_{f}J_{c}J}_{0\,\,0\,\,0}}{N_{J_{f}}^{BCS}N_{J_{c}}^{(g)}}\right)^{2}.
\label{Normmin2}
\end{equation}
The summations in Eqs. (\ref{Psi1}) and (\ref{Normmin2}) are restricted to the ranges 
$J_{f}\leq 24$  (see Table I), and  $J_{c}\leq 60$. In this way one accounts for all possible configurations of a given total angular momentum $J$, with $J$ running from 0 to 36.  The set of wave functions (\ref{Psi1}) describes the ground band which is associated to the case when all particles from the intruder orbital are paired. 

The {\it S}-band which crosses the {\it g}-band and produces the backbending in the energy spectra can be described by a  wave function similar to (\ref{Psi1}) with the difference that now one pair of particles is broken, i.e. they occupy two states which are not related by a time reversal transformation. The symmetry breaking is simulated by applying the angular momentum raising operator on a function with good symmetry. Thus the  2$qp$ state which is responsible for generating the 
$S$ band is a $K=1$ state of the following form:  
$J_+\alpha_{jk}^{\dagger}\alpha_{j-k}^{\dagger}|BCS\rangle_d$, where $\alpha^{\dagger}_{jk}$ is the creation quasiparticle operator  defined by the canonical BV transformation:
\begin{eqnarray}
\alpha_{jk}^{\dagger}&=&U_{jk}c_{jk}^{\dagger}-V_{jk}(-)^{j-k}c_{j-k},\nonumber\\
\alpha_{jk}&=&U_{jk}c_{jk}-V_{jk}(-)^{j-k}c_{j-k}^{\dagger}.
\label{BVtransf}
\end{eqnarray}
 The total projected state corresponding to the {\it S} band is defined by
\begin{equation}
\Psi_{JM;1}^{(2)}(jk)=\mathcal{N}_{J1}^{(2)}(jk)P_{M1}^{J}\left[J_+\alpha_{jk}^{\dagger}\alpha_{j-k}^{\dagger}|BCS\rangle_{d}\right]\psi_{g}=\mathcal{N}_{J1}^{(2)}(jk)\sum_{J_{f}J_{c}}\frac{C^{J_{f}J_{c}J}_{1\,\,0\,\,1}}{\mathcal{N}_{J_{f}1}^{jk}N_{J_{c}}^{(g)}}\left[\Phi_{J_{f}1}^{jk}\phi_{J_{c}}^{(g)}\right]_{JM},
\label{Psi2}
\end{equation}
where the normalization factor is
\begin{equation}
\left(\mathcal{N}_{J1}^{(2)}(jk)\right)^{-2}=\sum_{J_{f}J_{c}}\left(\frac{C^{J_{f}J_{c}J}_{1\,\,0\,\,1}}{\mathcal{N}_{J_{f}1}^{jk}N_{J_{c}}^{(g)}}\right)^{2}.
\end{equation}
The projected  two quasiparticle state $\Phi_{J_{f}1}^{jk}$ from (\ref{Psi2}) has the following expression:
\begin{equation}
\Phi_{J_{f}1;M_{f}}^{jk}=\mathcal{N}_{J_{f}1}^{jk}P_{M_{f}1}^{J_{f}}J_+\alpha_{jk}^{\dagger}\alpha_{j-k}^{\dagger}|BCS\rangle_{d}.
\label{phi2}
\end{equation}
The normalization factor of this state is defined by the matrix element
\begin{equation}
\left(\mathcal{N}_{J_{f}1}^{jk}\right)^{-2}=~_{d}\langle BCS|\alpha_{jk}\alpha_{j-k}J_-P_{11}^{J_{f}}J_+\alpha_{jk}^{\dagger}\alpha_{j-k}^{\dagger}|BCS\rangle_{d},
\end{equation}
which is determined in a similar way as the norm of the projected BCS function (see Appendix B).  The upper limit of the $J_{f}$ is still 24 even after the application of the quasiparticle creation operators, because it is the maximum angular momentum which can be realized in the $i_{13/2}$ intruder orbital, irrespective of the number of particles involved.

Both projected states, (\ref{Psi1}) and (\ref{Psi2}), depend on the deformation parameter $d$, although they are states with good angular momentum.   Note that the projected $2qp$ states are defined only for even angular momentum $J$,  with $J>1$. Another important property of the projected $2qp$ states is that two states with $k\neq k^{\prime}$ are not orthogonal.

\renewcommand{\theequation}{3.\arabic{equation}}
\section{The model Hamiltonian}
\label{sec: level3}

The backbending features of some rare earth nuclei will be studied with the following particle-core Hamiltonian:
\begin{equation}
H=H_{c}+H_{p}+H_{pair}+H_{pc}.
\end{equation}
The Hamiltonian $H_{c}$ is a harmonic  quadrupole boson operator:
\begin{equation}
H_{c}=\hbar\omega_{b}\sum_{\mu}b_{2\mu}^{\dagger}b_{2\mu},
\end{equation}
and describes a spherical core. $H_{p}$ is describing a set of particles in a spherical shell model single $j$ orbital of intruder nature, interacting through a paring force:
\begin{eqnarray}
H_{p}&=&(\varepsilon_{nlj}-\lambda)\sum_{m=all}c_{nljm}^{\dagger}c_{nljm},\nonumber\\
H_{pair}&=&-\frac{G}{4}P_{j}^{\dagger}P_{j},
\end{eqnarray}
where $P_{j}^{\dagger}(P_{j})$ are creation(annihilation) operators of the Cooper pair in the intruder orbital $j$ defined by
\begin{equation}
P^{\dagger}_{j}=\sum_{m>0}c_{nljm}^{\dagger}c_{nlj-m}^{\dagger}(-)^{j-m}.
\end{equation}
The operators $c_{nljm}^{\dagger}$ and $c_{nljm}$ stand for the creation and annihilation operators for a particle in the spherical shell model state $|nljm\rangle$ having the energy 
$\varepsilon_{nlj}$. The Lagrange multiplier $\lambda$ plays the role of the Fermi  energy for the paired system. For the sake of saving the space in what follows the spherical shell model state will be specified only by two quantum numbers, that is $|jm\rangle$.

The particle-core interaction is taken of the form:
\begin{eqnarray}
H_{pc}&\equiv& H_{qQ}+H_{J_fJ_c}\nonumber\\
&=&-A_{C}\sum_{\mu,m,m'}q_{2\mu}(j;mm^{\prime})c^{\dagger}_{jm}c_{jm'}\left((-)^{\mu}b^{\dagger}_{2-\mu}+b_{2\mu}\right)
+C\vec{J}_{c}\cdot\vec{J}_{f},\nonumber\\
q_{2\mu}(j;mm^{\prime})&=&\langle jm|r^{2}Y_{2\mu}|jm'\rangle ,
\end{eqnarray}
where $A_{C}$ and $C$ are  free parameters.  The last term plays an important role in reproducing the correct transition at the band crossing point and simulates the effect of the Coriolis coupling \cite{Bohr}.

Depending on whether the considered system is a near spherical or a deformed nucleus we diagonalize first the single-particle  plus the pairing interaction  and then treat the remaining terms or we diagonalize first the single-particle plus the $qQ$ interaction and then treat the rest in the resulting single-particle basis \cite{Ikeda}. Actually, here we make the option for the second procedure for reasons which will become clear in the next subsection.

\subsection{Pairing in a deformed single-particle basis}

The particle-core interaction leads to deforming the  single-particle mean-field. On the other hand
the interaction deforms the quadrupole boson. This mutual deformation effect is suggested for the following reasoning. If one considers the average of the particle-core Hamiltonian with a 
single-particle state one obtains a deformed Hamiltonian with a ground state described by an axially symmetric coherent state. As for the single-particle deformed mean-field let us consider, for simplification, the model Hamiltonian which corresponds to vanishing spin-spin interaction i.e., $C=0$ and unpaired particles ($G=0$),
\begin{equation}
\tilde{H}=H_p+H_c+H_{qQ}.
\end{equation}
Averaging this Hamiltonian with the coherent state
\begin{equation}
\psi_{b}=\exp{\left[d(b_{20}^{\dagger}-b_{20})\right]}|0\rangle,
\end{equation}
one obtains a single-particle Hamiltonian for a deformed mean-field similar to that used within the Nilsson model \cite{Nilsson}:
\begin{equation}
\tilde{H}_p=d^2\hbar\omega_b+H_p -2dA_C \sum_{m}q_{20}(j;m,-m)c^{\dagger}_{jm}c_{jm}.
\end{equation}
Note the role of nuclear deformation played by the parameter $d$.
Apart from an additive constant, in the first order of perturbation, the energies of 
$\tilde{H}_p$ are given by:

\begin{equation}
\varepsilon_{nljm}=\varepsilon_{nlj}-4dX_{C}(2n+3)C_{\frac{1}{2}\,2\,\frac{1}{2}}^{j\,\,2\,\,j}C_{m\,0\,m}^{j\,\,2\,\,j},
\end{equation}
where
\begin{equation}
X_{C}=\frac{\hbar}{8M\omega_{0}}\sqrt{\frac{5}{\pi}}A_{C}.
\end{equation}
Here $M$ and $\omega_{0}$ are the nucleon mass and the frequency of the harmonic oscillator function.
The dependence of single-particle energy on deformation is linear and shown in Fig.1 for the
$n=5$ shell.
 
We remark that in a single $j$ calculation, the $n$ and $l$ quantum numbers are superflue and therefore dropped out. Moreover, since only relative energies are involved in the BCS calculations the constant term, (i.e. that one not depending on $m$), is put equal to zero.
Therefore, in our calculations the single-particle space is spanned by $|jm\rangle$ and  the 
single-particle energies corresponding to the mentioned states are:

\begin{equation}
\varepsilon_{jm}=-4dX_{C}(2n+3)C_{\frac{1}{2}\,2\,\frac{1}{2}}^{j\,\,2\,\,j}C_{m\,0\,m}^{j\,\,2\,\,j},
\label{epsilonjm}
\end{equation}
with $n$ being the principal quantum number characterizing the major shell to which the intruder
state belongs.
The pairing interaction in such a deformed multiplet has been considered by B\'{e}s {\it at al} in Ref.
\cite{Bes}.
Note that the  deformation effect due to the $qQ$ interaction onto the single-particle state $|jm\rangle$, was ignored at this stage. According to Eq. (\ref{epsilonjm}) the energies of the time reversed states
are equal. Therefore, we can restrict the space to the states $|jm\rangle$ with $m>0$ keeping in mind that on each such state two nucleons are allowed.

\begin{figure}[h!]
\begin{center}
\includegraphics[width=1\textwidth]{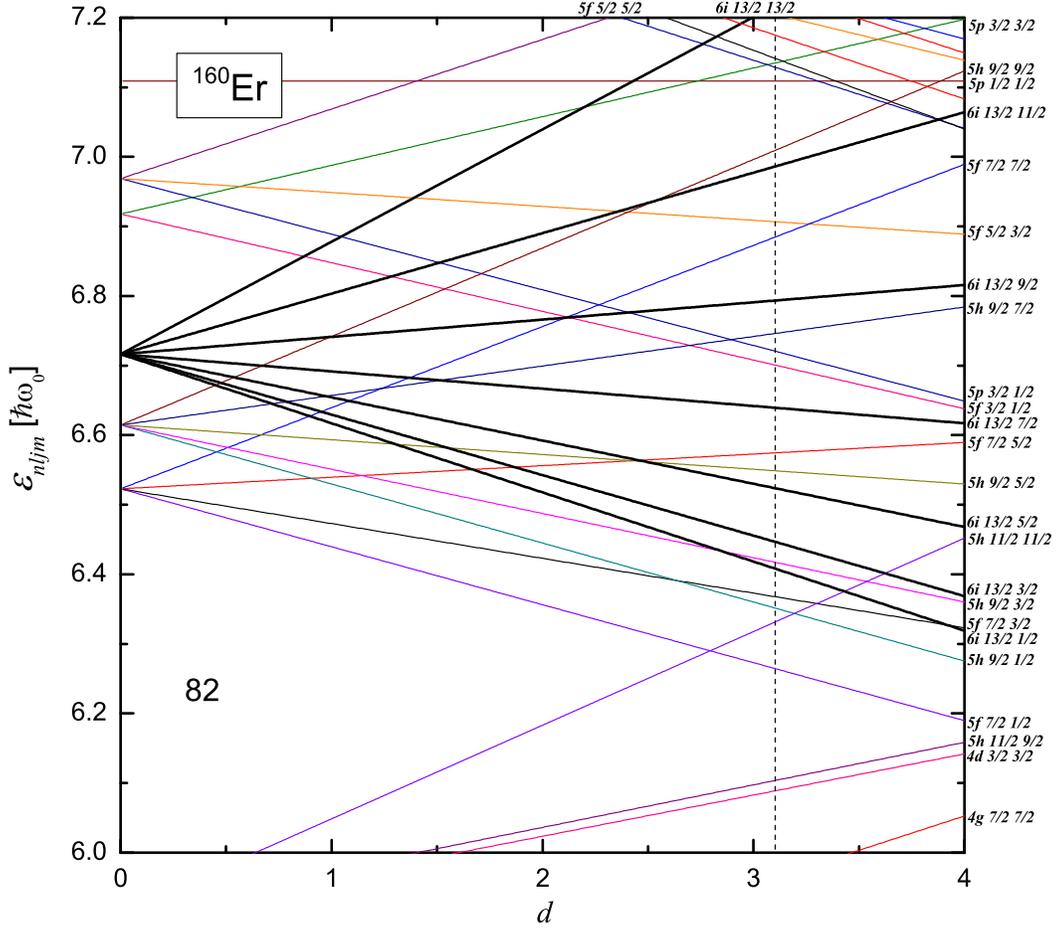}
\end{center}
\caption{(Color online) Neutron energies of the deformed single-particle states given by Eq. (3.9) with $X_C=50.81 ~keV$ are plotted as function of the deformation parameter $d$. The solid thick lines correspond to the $i_{13/2}$ states. Energies are given in units of $\hbar\omega_0$ 
($=41A^{-1/3}MeV$). The intersection of the dashed line, $d=3.103$, with the energy curves indicates the single-particle energies which correspond to $^{160}$Er. }
\label{Fig. 1}
\end{figure}
The single-particle Hamiltonian for the deformed mean-field can be written as:
\begin{equation}
H^{eff}_p=\sum_{m=all}\varepsilon_{jm}c^{\dagger}_{jm}c_{jm}.
\end{equation}

The next step is to treat $H^{eff}_p+H_{pair}$ through the BCS formalism, using the BV transformation (\ref{BVtransf}). In the quasiparticle representation the mentioned Hamiltonian becomes:
\begin{equation}
H_{qp}=E_{0}+\sum_{m=all}E_{jm}'\alpha_{jm}^{\dagger}\alpha_{jm}+\sum_{m>0}g_{jm}(-)^{j-m}(\alpha_{jm}^{\dagger}\alpha_{j-m}^{\dagger}+\alpha_{j-m}\alpha_{jm}),
\end{equation}
where the following notations were adopted:
\begin{equation}
E_{0}=-\lambda N_{part}-\frac{\Delta^{2}}{G},\,\,E_{jm}'=\frac{-\lambda(\varepsilon_{jm}-\lambda)+\Delta^{2}}{E_{jm}},\,\,g_{jm}=-\frac{\varepsilon_{jm}\Delta}{E_{jm}},\,\,\,\,(\varepsilon_{j}=0).
\end{equation}
Here $N_{part}$ represents the number of particles in the intruder orbital; $E_{jm}$ is quasiparticle energy  while $\Delta$ is the energy gap.

The expression of the $qQ$ interaction term in the $qp$ representation is:
\begin{eqnarray}
H_{qQ}&=&\left[2\sum_{m>0}q_{20}(j;mm)V_{jm}^{2}+\sum_{m=all}q_{20}(j;mm)(U_{jm}^{2}-V_{jm}^{2})\alpha_{jm}^{\dagger}\alpha_{jm}\right.\nonumber\\
&&\left.-2\sum_{m>0}q_{20}(j;mm)U_{jm}V_{jm}(-)^{j-m}(\alpha_{jm}^{\dagger}\alpha_{j,-m}^{\dagger}+\alpha_{j,-m}\alpha_{jm})
\right]\left(b^{\dagger}_{20}+b_{20}\right).
\end{eqnarray}
In deriving this expression we took into account the fact that only the component with
$\mu=0$ of the boson factor contributes when the average on the projected coherent state is calculated.

\subsection{The diagonalization of $H$ in the  particle-core product basis}
We recall the fact that the BCS  equations can be obtained either by minimizing the ground state energy or by vanishing the dangerous graphs. Indeed one can check that the total coefficient multiplying the cross terms
$(-)^{j-m}(\alpha_{jm}^{\dagger}\alpha_{j,-m}^{\dagger}+\alpha_{j,-m}\alpha_{jm})$,
coming from $H_{qp}$ and $H_{qQ}$ is vanishing.
 
The diagonal matrix elements of the various terms of the model Hamiltonian can be easily calculated. Their compact expressions are: 
\begin{eqnarray}
\langle\Psi_{JM}^{(1)}|H_{qp}+H_{qQ}|\Psi_{JM}^{(1)}\rangle&=&E_{0}+2dA_{eff},\\
\langle\Psi_{JM;1}^{(2)}(jk)|H_{qp}+H_{qQ}|\Psi_{JM;1}^{(2)}(jk)\rangle&=&E_{0}+2E_{jk}+2dA_{eff},
\end{eqnarray}
where
\begin{equation}
A_{eff}=-2X_{C}(2n+3)C^{j\,\,2\,\,j}_{\frac{1}{2}\,0\,\frac{1}{2}}\sum_{m>0}C^{j\,\,2\,\,j}_{m0m}V_{jm}^{2},
\end{equation}
is a constant quantity and can be omitted along with $E_{0}$.

The matrix elements of the harmonic boson Hamiltonian (3.2) on projected coherent states were given in Ref.\cite{Raduta2}, where a description for the ground band energies of axially deformed nuclei was provided. Thus, the  matrix elements of $H_{c}$ in the space of $0qp$ and $2qp$ states are given by:
\begin{eqnarray}
\langle\Psi_{JM}^{(1)}|H_{c}|\Psi_{JM}^{(1)}\rangle&=&\hbar\omega_{b}d^{2}\left(\mathcal{N}_{J}^{(1)}\right)^{2}\sum_{J_{f}J_{c}}\left(\frac{C^{J_{f}J_{c}J}_{0\,\,0\,\,0}}{N_{J_{f}}^{BCS}N_{J_{c}}^{(g)}}\right)^{2}\frac{I_{J_{c}}^{(1)}}{I_{J_{c}}^{(0)}},\\
\langle\Psi_{JM;1}^{(2)}(jk)|H_{c}|\Psi_{JM;1}^{(2)}(jk)\rangle&=&\hbar\omega_{b}d^{2}\left(\mathcal{N}_{J1}^{(2)}(jk)\right)^{2}\sum_{J_{f}J_{c}}\left(\frac{C^{J_{f}J_{c}J}_{1\,\,0\,\,1}}
{{\cal N}_{J_{f}1}^{jk}N_{J_{c}}^{(g)}}\right)^{2}\frac{I_{J_{c}}^{(1)}}{I_{J_{c}}^{(0)}}.
\end{eqnarray}

Using the tensorial form of $0qp$ and $2qp$ states and (\ref{Psi1}), (\ref{Psi2}) as well as the results of Appendix B, the matrix elements of the spin-spin interaction term $H_{J_fJ_c}$ are easily found:
\begin{eqnarray}
\langle\Psi_{JM}^{(1)}|H_{J_fJ_c}|\Psi_{JM}^{(1)}\rangle&=&\frac{C}{2}J(J+1)-\frac{C}{2}\left(\mathcal{N}_{J}^{(1)}\right)^{2}\sum_{J_{f}J_{c}}\left(\frac{C^{J_{f}J_{c}J}_{0\,\,0\,\,0}}{N_{J_{f}}^{BCS}N_{J_{c}}^{(g)}}\right)^{2}\nonumber\\
&\times&\left[J_{c}(J_{c}+1)+J_{f}(J_{f}+1)\right],\\
\langle\Psi_{JM;1}^{(2)}(jk)|H_{J_fJ_c}|\Psi_{JM;1}^{(2)}(jk)\rangle&=&\frac{C}{2}J(J+1)-\frac{C}{2}\left(\mathcal{N}_{J1}^{(2)}(jk)\right)^{2}\sum_{J_{f}J_{c}}\left(\frac{C^{J_{f}J_{c}J}_{1\,\,0\,\,1}}{{\cal N}_{J_{f}1}^{jk}N_{J_{c}}^{(g)}}\right)^{2}\nonumber\\
&\times&\left[J_{c}(J_{c}+1)+J_{f}(J_{f}+1)\right].
\end{eqnarray}
In this way the rotational spectra of {\it g}-band and {\it S}-band are completely determined.

\renewcommand{\theequation}{4.\arabic{equation}}
\section{Band hybridization procedure}
\label{sec: level4}

To describe the backbending phenomenon as being determined by the bands intersection we have to diagonalize the total Hamiltonian in a basis defined by (\ref{Psi1}) and (\ref{Psi2}). Unfortunately, this basis is not  orthogonal. Indeed the overlap matrix elements
\begin{equation}
\mathcal{O}_{12}^{JM}(jk)=\langle\Psi_{JM}^{(1)}(jk)|\Psi_{JM;1}^{(2)}(jk)\rangle,
\end{equation}
are not vanishing. This can be seen from the explicit expression:
\begin{equation}
\langle\Psi_{JM}^{(1)}|\Psi_{JM;1}^{(2)}(jk)\rangle=\mathcal{N}_{J}^{(1)}\mathcal{N}_{J1}^{(2)}(jk)
\sum_{J_{f}J_{c}}\frac{C^{J_{f}J_{c}J}_{0\,\,0\,\,0} C^{J_{f}J_{c}J}_{1\,\,0\,\,1}}{{N_{J_{c}}^{(g)}}^2
N^{BCS}_{J_f}{\cal N}^{jk}_{J_f1}}.
\end{equation}
Overlaps are nonvanishing due to the fact  that quasiparticle operators are not tensors of a definite rank and projection, which can be easily seen from their transformation against an arbitrary rotation  $R$:
\begin{eqnarray}
R\alpha_{jk}^{\dagger}R^{-1}&=&\sum_{m}\tilde{D}_{mk}^{j}\alpha_{jm}^{\dagger}+\tilde{\tilde{D}}_{mk}^{j}(-)^{j-m}\alpha_{j-m},\\
\tilde{D}_{mk}^{j}&=&(U_{jk}U_{jm}+V_{jk}V_{jm})D_{mk}^{j},\nonumber\\
\tilde{\tilde{D}}_{mk}^{j}&=&(U_{jk}V_{jm}-V_{jk}U_{jm})D_{mk}^{j}.\nonumber
\end{eqnarray}
That would not happen if the $U$ and $V$ coefficients were not dependent on the $m$ quantum number.
By diagonalizing  the overlap matrix, one obtains the eigenvalues $\alpha_{m}(J,jk)$ and the corresponding eigenvectors $V_{n}^{(m)}(J,jk)$. Then the functions
\begin{equation}
\Phi_{m}^{JM}(jk)=\left[\alpha_{m}(J,jk)\right]^{-1/2}\left[\Psi_{JM}^{(1)}V_{1}^{(m)}(J,jk)+
\Psi_{JM;1}^{(2)}V_{2}^{(m)}(J1,jk)\right],\,\,m=1,2,
\end{equation}
are mutually orthogonal and can be used as a  diagonalization basis for the total Hamiltonian 
\cite{Raduta3}. If we take the total wave function of the form
\begin{equation}
\Phi_{Tot}^{JM}(jk)=\sum_{m}X_{m}^{JM}(jk)\Phi_{m}^{JM}(jk),
\end{equation}
then we have to solve the following eigenvalue problem
\begin{equation}
\sum_{m'}\tilde{H}_{mm'}X_{m'}^{JM}(jk)=E_{JM}(jk)X_{m}^{JM}(jk),
\label{eigen}
\end{equation}
for finding the yrast spectrum. The Hamiltonian matrix $\tilde{H}_{nm}$ is defined by
\begin{equation}
\tilde{H}_{mm'}=\left[\alpha_{m}(J,jk)\alpha_{m'}(J,jk)\right]^{-1/2}\sum_{nn'}V_{n}^{(m)}(J,jk)\langle\Psi_{\beta}^{(n)}|H|\Psi_{\beta^{'}}^{(n')}\rangle V_{n'}^{(m')}(J,jk),
\end{equation}
where the low indices $\beta,\beta^{'}$ are either $JM$ if the corresponding upper index $n$ (or $n^{'}$) is $1$ or $JM;1$ for $n$ (or 
$n^{'}$) equal to 2.
By mixing the nonorthogonal states in the orthogonalization process and then diagonalizing the model Hamiltonian a natural interaction of the primary bands takes place.
Such an interaction is angular momentum dependent and seems to be more efficient in the $K=1$ band hybridization process \cite{Bonatsos} than a constant one.

The off-diagonal  matrix elements of the model Hamiltonian  terms, in the non-orthogonal basis are given by the following expressions:
\begin{eqnarray}
\langle\Psi_{JM}^{(1)}|H_{qp}+H_{qQ}|\Psi_{JM;1}^{(2)}(jk)\rangle&=&\mathcal{N}_{J}^{(1)}\mathcal{N}_{J1}^{(2)}(jk)
\sum_{J_{f}J_{c}}\frac{C^{J_{f}J_{c}J}_{0\,\,0\,\,0} C^{J_{f}J_{c}J}_{1\,\,0\,\,1}}{{N_{J_{c}}^{(g)}}^2N^{BCS}_{J_f}{\cal N}^{jk}_{J_f1}}\nonumber\\
&\times&\left(E_{0}+E_{jk}+2dA_{eff}\right),\\
\langle\Psi_{JM}^{(1)}|H_{c}|\Psi_{JM;1}^{(2)}(jk)\rangle&=&\hbar\omega_{b}d^{2}
\mathcal{N}_{J}^{(1)}\mathcal{N}_{J1}^{(2)}(jk)
\sum_{J_{f}J_{c}}\frac{C^{J_{f}J_{c}J}_{0\,\,0\,\,0} C^{J_{f}J_{c}J}_{1\,\,0\,\,1}}{{N_{J_{c}}^{(g)}}^2N^{BCS}_{J_f}
{\cal N}^{jk}_{J_f1}}\frac{I_{J_{c}}^{(1)}}{I_{J_{c}}^{(0)}},\\
\langle\Psi_{JM}^{(1)}|H_{J_fJ_c}|\Psi_{JM;1}^{(2)}(jk)\rangle&=&\frac{C}{2}
\mathcal{N}_{J}^{(1)}\mathcal{N}_{J1}^{(2)}(jk)
\sum_{J_{f}J_{c}}\frac{C^{J_{f}J_{c}J}_{0\,\,0\,\,0} C^{J_{f}J_{c}J}_{1\,\,0\,\,1}}{{N_{J_{c}}^{(g)}}^2N^{BCS}_{J_f}{\cal N}^{jk}_{J_f1}}\nonumber\\
&\times&(J(J+1)-J_{c}(J_{c}+1)-J_{f}(J_{f}+1)).
\end{eqnarray}
\renewcommand{\theequation}{5.\arabic{equation}}
\section{Numerical application}
\label{sec: level5}

The formalism described in the previous sections was applied to six even-even nuclei from the rare earth region which are known to be good backbenders, namely $^{156}$Dy, $^{160}$Yb, $^{158,160}$Er and $^{164,166}$Hf. The first three nuclei are $N=90$ isotons, the next two are $N=92$ isotons, and the last one has 94 neutrons. 

The model Hamiltonian involves four free parameters namely, the pairing constant $G$, quadrupole-quadrupole interaction strength $X_{C}$, the spin-spin interaction strength $C$ and the boson frequency of the core $\hbar\omega_{b}$. Another parameter is the deformation parameter $d$, defining the coherent state $\psi_b$.

\begin{table}[h!]
\caption{The pairing strength ($G$) is given in units of MeV, while the  quadrupole-quadrupole ($X_C$) and spin-spin ($C$) interactions strength are given in units of keV. The boson frequency of the core is also given. The list of the deformation parameters $d$ are presented together with the corresponding $\beta_{2}$ deformation, taken  from Ref.\cite{Lalazissis}. The manner in which these parameters were fixed is explained in the text.}
\begin{tabular}{|c|c|c|c|c|c|c|}
\hline
~~~~Nucleus~~~~&~~~~~~$\beta_{2}$~~~~~~&~~~~~~$d$~~~~~~&~~~$\hbar\omega_{b}$(MeV)~~~&~~~$X_{C}
$(keV)~~~&~~~$G$(MeV)~~~&~~~$C$(keV)~~~\\
\hline
$^{156}$Dy&0.211&3.1414&0.9781&52.63&0.1786&7.227\\
$^{160}$Yb&0.195&2.5000&1.0078&65.59&0.1854&3.690\\
$^{158}$Er&0.203&2.7355&0.9778&59.95&0.1814&6.010\\
$^{160}$Er&0.231&3.1030&1.0089&50.81&0.1769&9.363\\
$^{164}$Hf&0.208&2.6605&1.1039&59.89&0.1836&5.220\\
$^{166}$Hf&0.237&2.8505&1.0436&52.89&0.1737&6.963\\
\hline 
\end{tabular}
\end{table}  
 The input data for the BCS equations are the pairing constant $G$ and the single-particle energies determined by Eq.(3.9).
 The deformation parameter $d$ and the boson frequency $\hbar\omega_{b}$ are fixed so  that the first energy levels lying before the band crossing point are reproduced with a good accuracy by  Eq.(3.19). Indeed, the $g$ band energies are not very sensitive to the single-particle degrees of freedom. Practically, the only  contribution to the total energy is due to the core, 
the particle-core interaction being reflected in the norm of the projected BCS state. In the case of the $g$ band states, all particles are paired, and  the dominant term of the sums in Eqs.(3.19) and (3.21) corresponds to the situation $J_{f}=0$ and $J=J_{c}$. This is also consistent with the fact that until the band crossing point the whole angular momentum is carried by the core. In this way the $g$band energies can be roughly approximated by:
\begin{equation}
E_{J}\approx \hbar\omega_{b}d^{2}\frac{I_{J}^{(1)}}{I_{J}^{(0)}},
\end{equation}
which is just the expression of the ground band energies predicted by the coherent state model \cite{Raduta0,Raduta1}. In the extreme limits of large and small deformations $d$, the above energies have been approximated by compact and simple functions of $J(J+1)$ in Refs.\cite{Raduta4,Raduta5}. Actually, this simple expression can be used to determine the parameters $d$ and $\hbar\omega_{b}$ by fitting the first energies before the band crossing of the $g$ and $S$ bands. Later on a tuning fit can be achieved by using the eigenvalues of the model Hamiltonian in the orthogonal basis.

The pairing interaction constant $G$ and the $qQ$ interaction strength  $X_{C}$ are fixed such that the band crossing point and the observed sequence of single-particle energies are reproduced. Here we deal with neutrons  from the shell $i_{13/2}$, where we can expect at most seven pairs.  The number of neutrons considered out of the core, in the shell
$i_{13/2}$, is two for $^{156}$Dy, $^{158}$Er, $^{160}$Yb, $^{164}$Hf,$^{160}$Er and four for $^{166}$Hf. Solving the BCS equations one obtains the gap parameter $\Delta$,
the Fermi level energy $\lambda$ and consequently the occupation probability parameters $U$ and $V$. With this data the wave functions corresponding to $0qp$ (\ref{Psi1}) and $2qp$ (\ref{Psi2}) are completely determined. The relevant information yielded by the BCS calculations are presented in Table III. We mention the fact that our choice for the number of neutrons distributed on the 
$i_{13/2}$ substates is consistent with the BCS calculations in an extended single-particle space.
Indeed, we solved the BCS equations for a space of 23 states, consisting in the union of the 
substates $5h_{9/2m}, 5f_{7/2m},6i_{13/2m},5p_{3/2m},5f_{5/2m},5p_{1/2m},5h_{11/2,11/2}$, where we distributed
10,12 and 14 neutrons for the $N=90,92,94$ isotons respectively. The results for the fitted pairing strengths can be interpolated by the following function:
\begin{equation}
G=\frac{1}{A}\left(g_0-g_1\frac{N-Z}{A}\right), \rm{with} ~g_0=37.44 MeV,~ g_1=62.25 MeV.
\end{equation}
Summing up the occupation probabilities for the $i_{13/2}$ sub-states, we obtained the average number $\langle N^{i_{13/2}}_{part}\rangle $ listed in Table III. Thus $N_{part}$ is the largest even number smaller than $\langle N^{i_{13/2}}_{part}\rangle $. The exception is for the case of $^{166}$Hf where $N_{part}$ is larger than, but very close to, $\langle N^{i_{13/2}}_{part}\rangle $.
The calculations in the reduced space of $i_{13/2}$ substates was performed with a pairing strength chosen such that the minimal $qp$ energy is equal to that obtained within the extended space. In this way the occupation probabilities obtained by solving the BCS calculations in the extended and reduced single-particle space, respectively, are close to each other. {\it In his context one could say that the schematic calculations in the reduced single-particle space accounts for the effective pairing interaction in this subspace.} 

\begin{table}[h!]
\caption{The Fermi level energies, gap parameters and the quasiparticle energies are given for the chosen number of particles $N_{part}$, and the projection $k$ associated with the broken pair. We also give the average number of particles, denoted by $\langle N^{i_{13/2}}_{part}\rangle$, in the multiplet $i_{13/2}$.}
\begin{tabular}{|c|c|c|c|c|c|c|c|}
\hline
~Nucleus~&~~~$N$~~~&~$N^{i_{13/2}}_{part}$~&~$\langle N^{i_{13/2}}_{part}\rangle$~&~~~$\lambda$(MeV)~~~&~~~$\Delta$(MeV)~~~&~$k(E_{k}=min)$~&~~~$E_{k}$(MeV)~~~\\
\hline
$^{156}$Dy&90&2&2.39&48.7643&1.12446&1/2&1.12448\\
$^{160}$Yb&90&2&2.47&48.3187&1.19819&1/2&1.19830\\
$^{158}$Er&90&2&2.44&48.5354&1.15443&1/2&1.15460\\
$^{160}$Er&92&2&3.12&48.6732&1.16028&3/2&1.16036\\
$^{164}$Hf&92&2&3.18&48.2444&1.23826&3/2&1.23827\\
$^{166}$Hf&94&4&3.88&48.4181&1.17327&3/2&1.20183\\
\hline
\end{tabular}
\end{table}

Note that except for $^{166}$Hf, the gap  $\Delta$ is very close in magnitude to the corresponding quasiparticle energies, which suggests that the Fermi level lies close to the selected $k$-energy level. Of course,  the  $S$ band associated to the $2qp$ projected state $(k)^2$ is the first one which intersects the $g$ band and will become the yrast    after intersection.
{\it The quantum numbers $k$ listed in Table III are consistent with the Nilsson model prediction for the last filled orbital of the chosen nuclei having the quadrupole nuclear deformation 
$\beta_2$ given in Table II.}

Our calculations show that the $J_{f}=0$ component has a squared weight of about 45\%-64\% in the BCS composition. 
 As a matter of fact, this suggests that the approximation (5.1) made for the {\it g}-band energies is to be corrected due to the nonvanishing angular momentum components of the BCS state.
In fact since the $J_{f}=0$ component is only moderately dominant in the BCS state composition,
the effective angular momentum of fermions is not vanishing in the $g$ band and not 12 for the states belonging to the $S$ band. This feature is conspicuous from the analysis of Fig. 5.

The matrix elements between states (2.18) of the spin-spin interaction term have a very peculiar 
feature. Their numerical values increase with the total angular momentum, going from negative values to positive ones. The transition from negative to positive values takes place at $J=10,12$, where, in most cases, the backbending shows up. In this situation the parameter $C$  does not affect the position of the band crossing point. However, it influences the magnitude of the yrast energies in the high spin region. Due to this feature  $C$ is fixed as to reproduce the moderate high spin energies of the yrast band. The fitted parameters of the model Hamiltonian are collected in Table II. 
We remark that the factor $dX_C$ which, according to Eq.(3.9), plays the role of the deformed mean-field strength, has roughly a linear dependence on the quadrupole nuclear deformation
$\beta_2$. This dependence is illustrated in Fig.2.

\begin{figure}[h!]
\begin{center}
\includegraphics[width=1\textwidth]{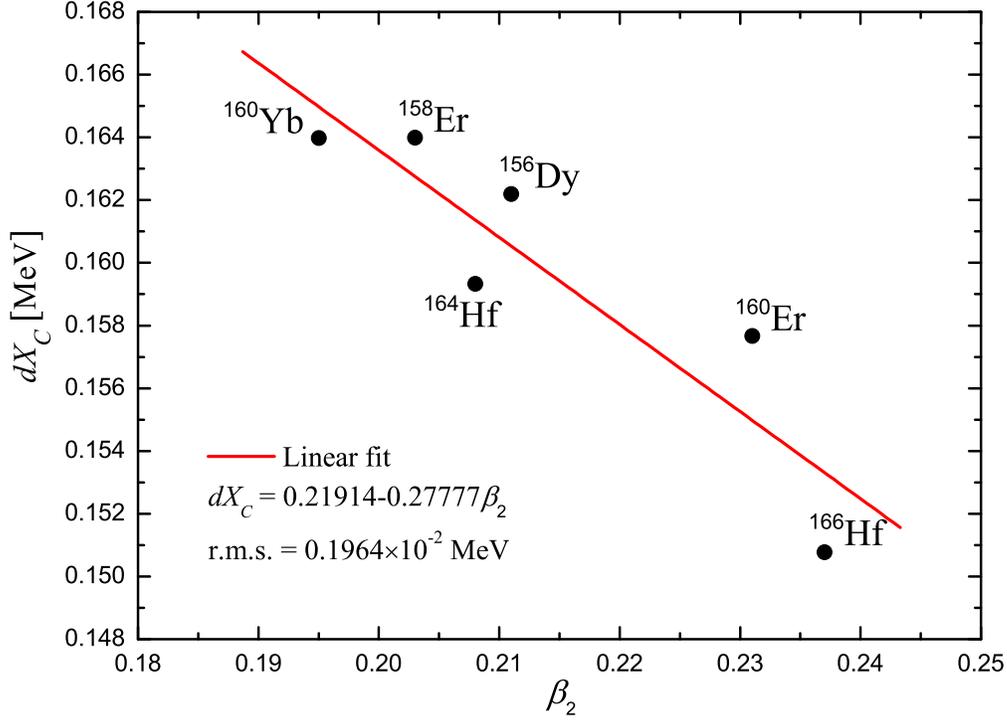}
\end{center}
\caption{(Color online) The product $dX_C$, with the factors given in Table II,
is represented as a function of the nuclear deformation $\beta_2$ taken from Ref.\cite{Lalazissis}.}
\label{Fig. 2}
\end{figure}
Using the parameters specified in Table II Eq. (\ref{eigen}) provides the system energies. For each angular momentum $J$ we select the lowest eigenvalues $E(J)$. The set of energies
$E_J$ defines the yrast band.
The discrete derivatives of the resulting energies with respect to the angular momentum defines
the angular frequency:
\begin{equation}
\hbar\omega(J)=\frac{dE(J)}{dJ}\approx\frac{1}{2}\left[E(J+2)-E(J)\right].
\end{equation}
Alternatively, the angular velocity can be also defined by using for $E(J)$ the expression 
provided by a symmetric rotor Hamiltonian:
\begin{equation}
E(J)=\frac{J(J+1)}{\mathfrak{I}}.
\end{equation}
Then the discrete derivative of this expression yields:
\begin{equation}
\hbar\omega(J)=\frac{2J+3}{\mathfrak{I}}.
\end{equation}
From here one derives a simple expression for the  moment of inertia:
\begin{equation}
\mathfrak{I}=\frac{4J+6}{E(J+2)-E(J)}.
\end{equation}
The backbending plot is a graph in which the moment of inertia is plotted versus $(\hbar\omega)^{2}$. Theoretical results and experimental data are usually compared in terms of  this plot. This is done in  Fig.3 for the even-even rare earth nuclei treated here. Note that in all cases the zigzag behavior is reproduced quite well.
In general, a good agreement between  theoretical and experimental data  corresponding to low-spin states and moderately high spin states located however below the possible second backbending
is obtained. The second backbending is known to be caused by the consecutive breaking of a neutron $i_{13/2}$ pair and a proton pair from the shell $h_{11/2}$ . The second backbending seems to be a rare event. Despite  this  some of the considered nuclei  exhibit  such a phenomenon. Indeed,
 $^{160}$Yb has a second backbending which occurs at $J$=26, while for $^{158}$Er the second backbending shows up in the same region but is less evident being rather an up-bending. In both mentioned cases our calculations show an up-bending around the J=26 state. The study of the second backbending is however beyond the scope of the present paper.

The crossing of the $g$ and the $S$ bands is illustrated in Fig.4, where their energies are plotted as a function of $J(J+1)$. We also give the results for the two unperturbed bands whose energies are approximated as the diagonal matrix elements of the system's Hamiltonian in the 
nonorthogonal basis. In each panel the $r.m.s.$ values for the deviation of the theoretical results from the corresponding experimental data are given.
 As can be seen from Fig.4, while  the {\it S}-band energies exhibit  a linear dependence  on $J(J+1)$ for the {\it g} band a slight quadratic dependence on $J(J+1)$ is met. Since $^{156}$Dy and
$^{160}$Er have the largest deformation $d$, the linear dependence on $J(J+1)$ of the corresponding
$g$ band energies prevail.

\begin{figure}[h!]
\begin{center}
\includegraphics[width=1\textwidth]{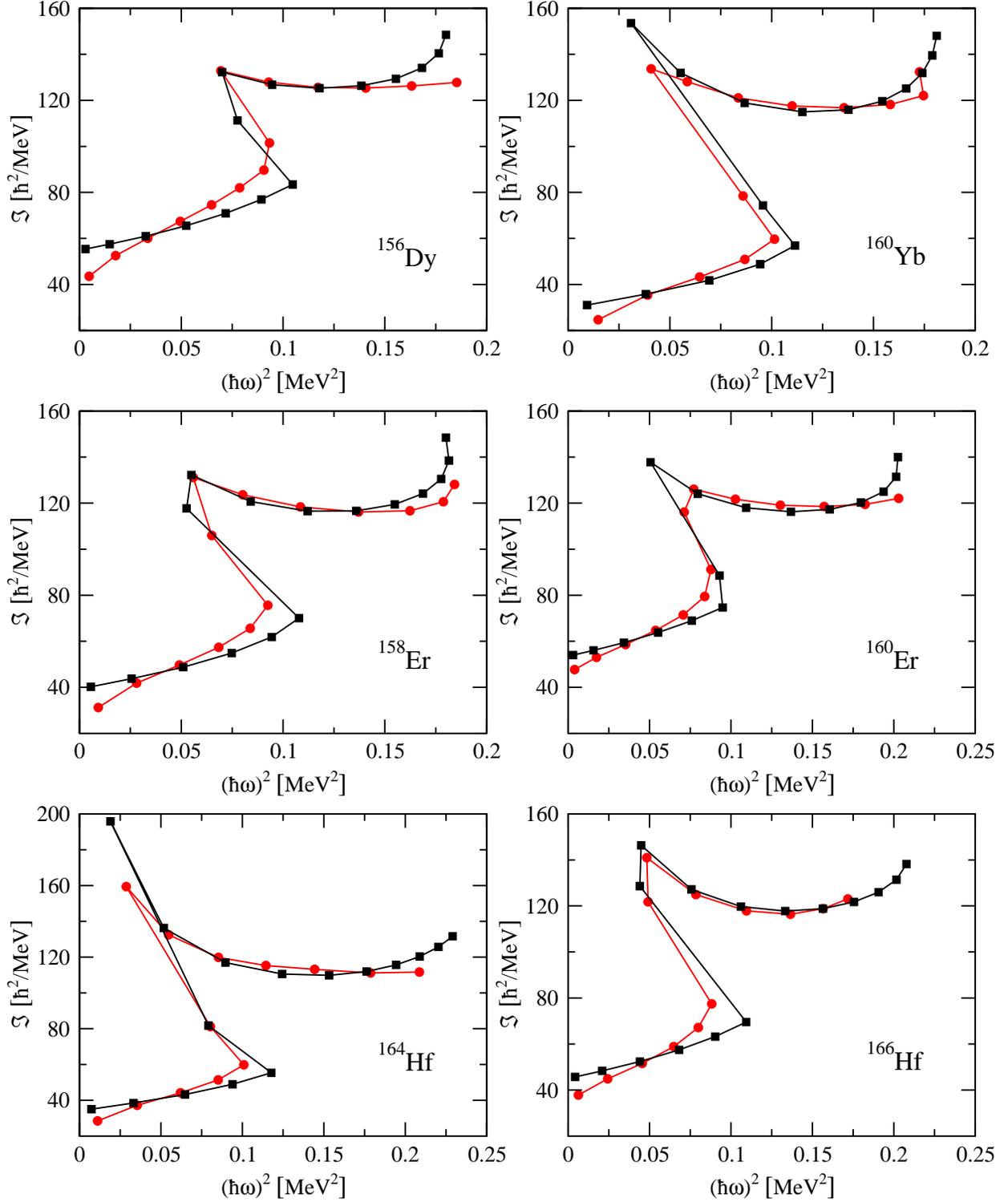}
\end{center}
\caption{(Color online) Backbending plots for $^{156}$Dy, $^{160}$Yb, $^{158,160}$Er and $^{164,166}$Hf isotopes comparing theory (black squares) with experiment (red circles). Experimental data are taken from \cite{Reich1,Reich2,Helmer,Balraj,Shurshikov}.}
\label{Fig. 3}
\end{figure}

\begin{figure}[h!]
\begin{center}
\includegraphics[width=0.85\textwidth]{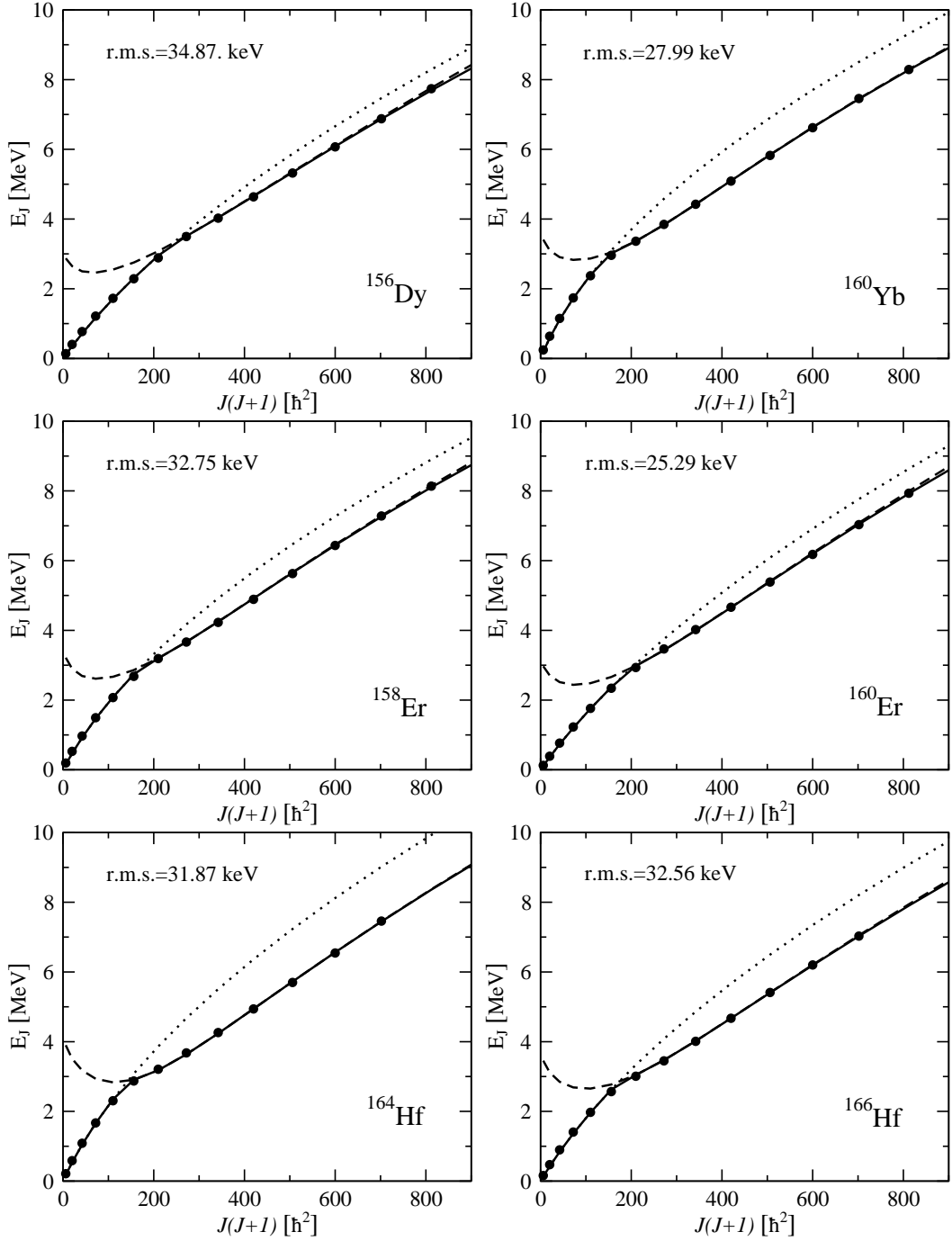}
\end{center}
\caption{Energy trajectories resulting from the diagonalization of total Hamiltonian in the orthogonal basis (4.4) (straight lines) are compared with experimental points (black circles) for $^{156}$Dy, $^{160}$Yb, $^{158,160}$Er and $^{164,166}$Hf. The unperturbed {\it g}-band (dotted lines) and the selected $2qp$ band (dashed lines) are also presented. For each nucleus the 
$r.m.s.$ values corresponding only to the states taken in consideration here, i.e. up to $J=26,28$, are given.}
\label{Fig. 4}
\end{figure}


In order to investigate the alignment of the individual angular momenta to the angular momentum of the core we define the average angular momenta  for the interacting components as follows: 
\begin{eqnarray}
\tilde{J}_{c}(\tilde{J}_{c}+1)&=&\langle\Phi_{Tot}^{JM}(jk)|\vec{J}_{c}^{2}|\Phi_{Tot}^{JM}(jk)\rangle,\\
\tilde{J}_{f}(\tilde{J}_{f}+1)&=&\langle\Phi_{Tot}^{JM}(jk)|\vec{J}_{f}^{2}|\Phi_{Tot}^{JM}(jk)\rangle.
\label{JfsiJc}
\end{eqnarray}
The full alignment corresponds to the situation when $\tilde{J}_{c}+\tilde{J}_{f}$ equates the total angular momentum of the system $J$. The departure from this ideal picture is measured by the deviation:
\begin{equation}
\Delta J=\left|J-(\tilde{J}_{c}+\tilde{J}_{f})\right|.
\end{equation}
This quantity together with the average angular momenta of the core  $\tilde{J}_{c}$ and fermion system $\tilde{J}_{f}$ are plotted versus the  total angular momentum $J$, in Fig. 5.  These plots reveal several features, such as the band crossing point, the amount of angular momentum carried by the broken pair or even the fraction of the angular momenta alignment. For example for Er isotopes and $^{166}$Hf the band crossing takes place at $J=12$, while for $^{164}$Hf  and $^{160}$Yb at $J=10$  and for $^{156}$Dy  at $J=14$. This is actually a confirmation of the backbending results from Fig.3. The amount of angular momentum, $\tilde{J}_{f}$, carried by the broken pair varies from 10$\hbar$ 
to 14$\hbar$. We remark that at the band crossing the alignment defect is maximal and this is decreasing by increasing $J$. The full alignment is never reached but the defect $\Delta J$ exhibits a plateau with the value of 2$\hbar$-3$\hbar$, beyond $J=20$.  This is a salient feature for the present formalism which is not met in other approaches
where $\Delta J_c =\Delta J_f=10,12$. Indeed, for $^{160}$Yb the variation of $J_c$ around crossing is only 2$\hbar$ and, moreover, after band crossing the core and fermionic angular momenta are almost equal to each other. For all other nuclei, at the crossing $J$
the core angular momentum variation is about 4$\hbar$. We don't have any situation where after crossing $J_c=0$ . Therefore for a critical value of angular momentum of the core the spin-spin interaction causes the de-pairing of two neutrons, which  almost align their angular momenta and starting from a larger total spin ($\sim 20$)  the fermion total angular momentum is almost aligned to the
core angular momentum, $\Delta J=2\hbar-3\hbar$. In general, the second alignment is produced for $J_c>J_f$.

\begin{figure}[h!]
\begin{center}
\includegraphics[width=1\textwidth]{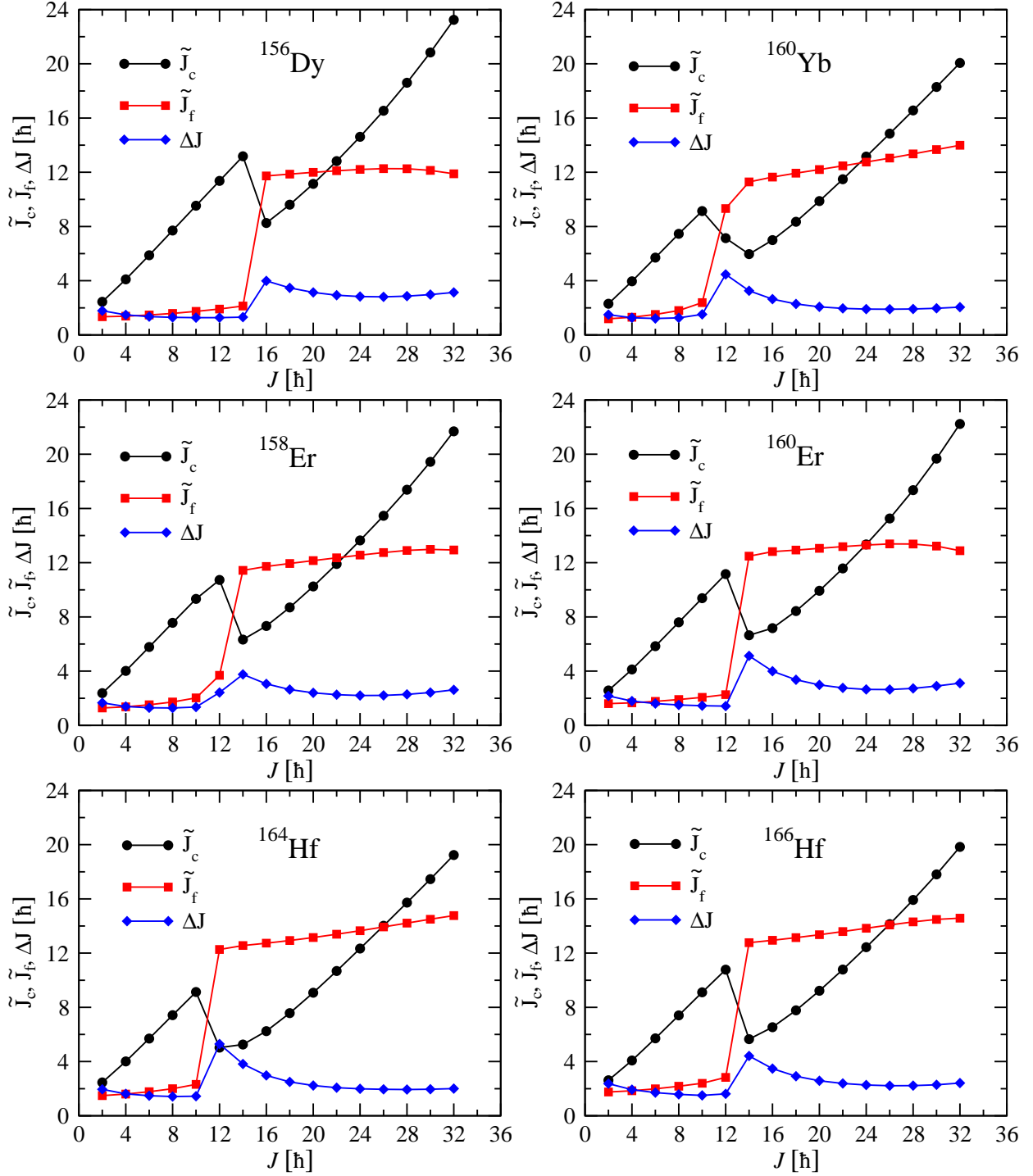}
\end{center}
\caption{(Color online) Expected angular momentum of the core and of the broken pair of intruder neutrons. The deviation of total angular momentum $\Delta J$ is also presented.}
\label{Fig.5}
\end{figure}

\begin{figure}[h!]
\begin{center}
\includegraphics[width=0.5\textwidth]{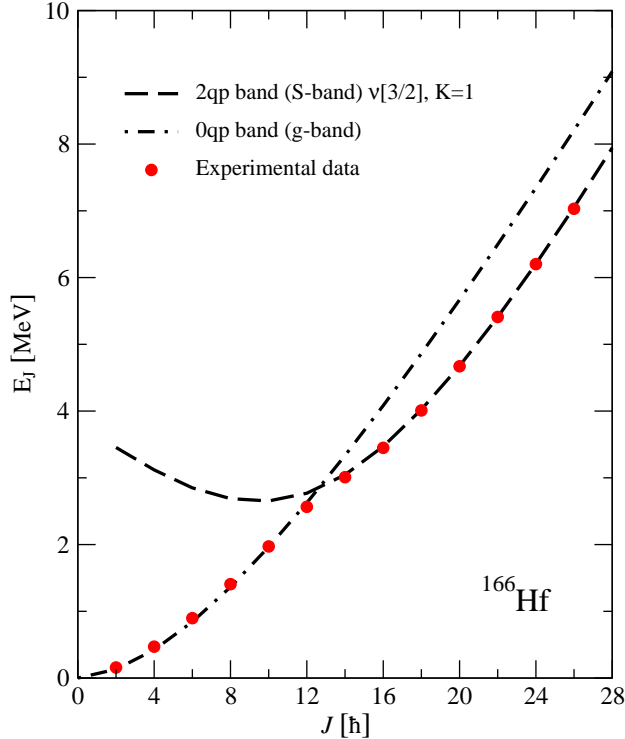}
\end{center}
\caption{(Color online) Energy plot for $^{166}$Hf showing the interacting bands. Rotational energy of the {\it g}-band and of the $2qp$ band from a certain spin increases monotonously. Experimental data points are also visualized.}
\label{Fig. 6}
\end{figure}

Figure 6 shows the calculated rotational energies  and the experimental data of {\it g}-band and {\it S}-band versus angular momentum for $^{166}$Hf. We present this case separately since here  the best agreement  of theoretical results and the corresponding data was obtained in the backbending region, after the crossing point. The slope of the curves from this figure is the angular velocity $\omega(J)$. Note that up to a critical spin the rotational energies of the $2qp$ band have a negative slope then this is vanishing and further
is increasing. Finally it reaches a constant value which reflects an equidistant structure for the spectrum in this region. The later situation shows up when a maximal alignment is achieved. The negative slope corresponds to the situation when the fermion and the core angular momenta make an angle varying from $\pi$ to $\pi/2$. At the beginning of the interval the individual angular momenta are almost anti-aligned and their sum is also almost antialigned to the core angular momenta. When the individual angular momenta are partially antialigned, the two frequencies caused by particles and core respectively, add destructively. Increasing $J$ the two frequencies have the same sign and therefore add each other, constructively. In this part of the curve, the second alignment namely that of the total fermionic and the core angular momenta, starts operating. When a maximal alignment is achieved the slope keeps constant and consequently the two curves become almost parallel. As shown in Fig.5, the properties described before, seem to be generally valid. 
A similar angular momentum dependency was obtained also in Refs.\cite{Stephens,Hara1}. 

The mechanism of breaking the pairs and aligning the individual angular momenta to the core angular momentum is suggested in Fig.7. Indeed, for low values of the core angular momentum the nucleon angular momenta are aligned to the symmetry axis of the mean-field which, as a matter of fact, is determined by the $qQ$ coupling term. This situation is shown in Fig.7 a) where 
$\vec{J}_f=0$ and $\vec{J}_c=\vec{J}$. 

The mechanism suggested in Fig.7 is consistent with the following phenomenological picture.
The core angular momentum is perpendicular to the symmetry axis OZ. Let us consider that OX is the direction of $\vec{J_c}$. Averaging $H_{J_fJ_c}$ with the Wigner functions $D^{J_c}_{MK}$ with
$M$ and $K$ being eigenvalues of $(J_c)_x$, one obtains an operator acting in the intrinsic frame
which breaks the time reversal symmetry. Due to this property this is the term responsible for the neutron pair breaking connecting the states of $0qp$ and $2qp$ types, respectively. In the intrinsic frame the mentioned term connects a state with all particles paired, (i.e. with total $K$ equal to zero) with a state where a $K=0$ pair is replaced by a $K=1$ broken pair. Actually, the same 
effect is obtained within microscopic  models using for a many-body system with pairing interaction a cranking term of the type $\omega j_x$. It is worth mentioning the following specific features of the present model. The total quantum number $K$ is obtained by summing the contributions from nucleons and the core. However, writing the projected state of the core [25] in the intrinsic frame, one obtains an expression which is a superposition of components of different $K$. Among these, the major component corresponds to $K=0$. In this respect the broken pair state is not a pure $K=1$ state but, due to the core, a superposition of various $K\ne 0$ components. However the major component is the one characterized by $K=1$.

At this stage it is worth making clear the way the spin-spin interaction simulates the action of the Coriolis force in the intrinsic reference frame. Indeed, up to a diagonal term the interaction $H_{J_fJ_c}$ can be written  in the following form:
\begin{equation}   
H_{J_fJ_c}\sim J_{f-}J_{+}+J_{f+}J_{-}
\end{equation}
with $J$ denoting the total angular momentum. Since the core projected state is a superposition of components  where the core function factor is a Wigner function of even values for  $K$,  the projected state characterizing the whole system is a superposition of components having a Wigner
function with odd $K$ as one of the factor states. The action of the raising or lowering operator on $D^J_{MK}$ will transform it to $D^J_{MK+1}$ or $D^J_{MK-1}$ with an even projection of $\vec{J}$ on the symmetry axis. The factor states associated with quasiparticles will be transformed from a $K=1$ state either to a $K=0$ state or to a $K=2$ state. Concluding, in the intrinsic frame the Coriolis term connects the $K=1$  2$qp$ states with the $K=0$  2$qp$ states. In the case of the odd particle-core system with one particle out of the core, the matrix element of the Coriolis interaction is different from zero only for the states with $K=\pm\frac{1}{2}$. The interaction is attractive or repulsive depending on whether $J+1/2$ is even or odd. 
Note that, by contrast, here a set of even numbers of nucleons is moving outside a phenomenological core and, moreover, the Coriolis interaction is effective in any $K$ state.
In the laboratory frame the spin-spin interaction affects mainly the $2qp$ states, as is shown in Fig.\ref{Fig.9}.
It is worth noticing that the interaction is attractive in the $2qp$ band states having an angular momentum smaller than the crossing point and repulsive in other states of the $S$ band. Therefore, the spin-spin interaction causes an attenuation effect on the moment of inertia in the $S$ band, after crossing the $g$ band.
If we switch off the spin-spin interaction, the slope of the curves from Fig.4, representing the energy versus $J(J+1)$, is decreased  after the two bands' crossing point, which results in enlarging the moment of inertia of the $S$ band. Due to this feature the bending in the moment of inertia plot is more pronounced than the bending taking place in the presence of the spin-spin interaction.

The alignment of the particle angular momenta to the core angular momenta
is shown in Fig.7 b). The fact that the total $K$  for the two neutrons is equal to unity prevents a full alignment. The larger $\vec{J}_f$ the larger  the effect of the spin-spin interaction term. This infers that the backbending occurs when the broken pair is from a large spin single-particle state. For this reason the most favorable candidates for decoupling are particles from intruder orbitals like $i_{13/2}$ for neutrons and $h_{11/2}$ for protons.

We stress on the fact that breaking a pair means to break the time reversal symmetry of the system, (i.e., to promote a paired particle to a state  of different $k$ which results in having a pair with a total $K$ different from zero). Such an operation cannot be performed by the $qQ$  interaction [see Eq.(3.15)] since the quasiparticle factor has only two quasiparticle  $K=0$ terms. Therefore in the present formalism the only term responsible for pair breaking is the spin-spin interaction. 

\begin{figure}[h!]
\begin{center}
\includegraphics[width=0.9\textwidth]{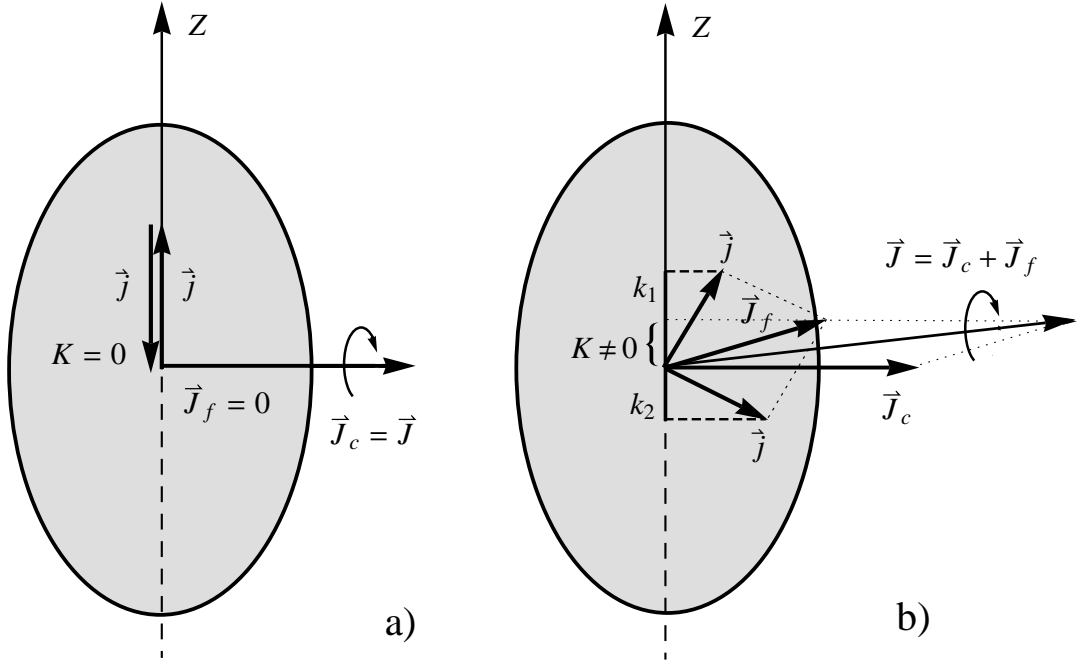}
\end{center}
\caption{(Color online) The coupling scheme for {\it g}-band (a) and for {\it S} band (b) associated with a broken pair of particles which give rise to a $\vec{J}_{f}$ angular momentum and a general $K=k_{1}+k_{2}$ projection. The angular momentum of the particles is then coupled to that of the prolate core which is perpendicular to the symmetry axis ($Z$), resulting in the total angular momentum $\vec{J}$.}
\label{Fig. 7}
\end{figure}
\begin{figure}[h!]
\begin{center}
\includegraphics[width=0.9\textwidth]{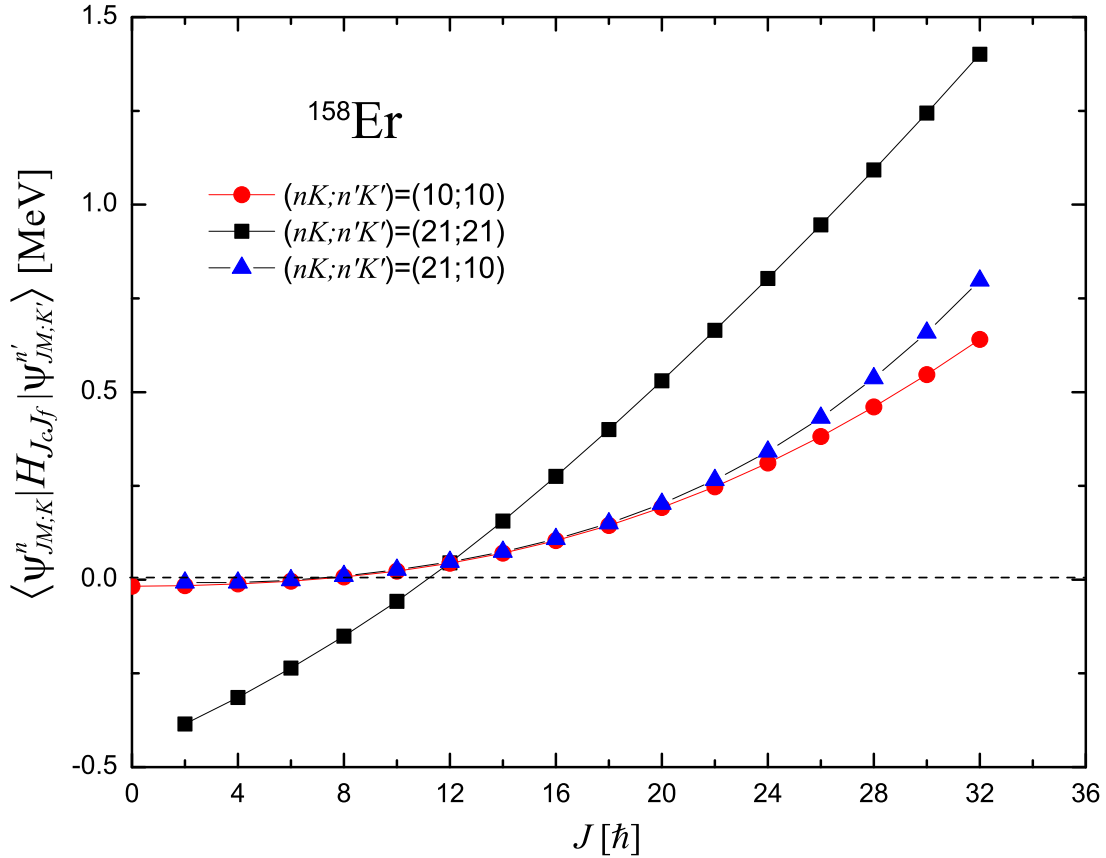}
\end{center}
\caption{(Color online) The matrix elements of the spin-spin interaction are plotted as function of the total angular momentum. }  
\label{Fig.9}
\end{figure}
\begin{figure}[h!]
\begin{center}
\includegraphics[width=0.9\textwidth]{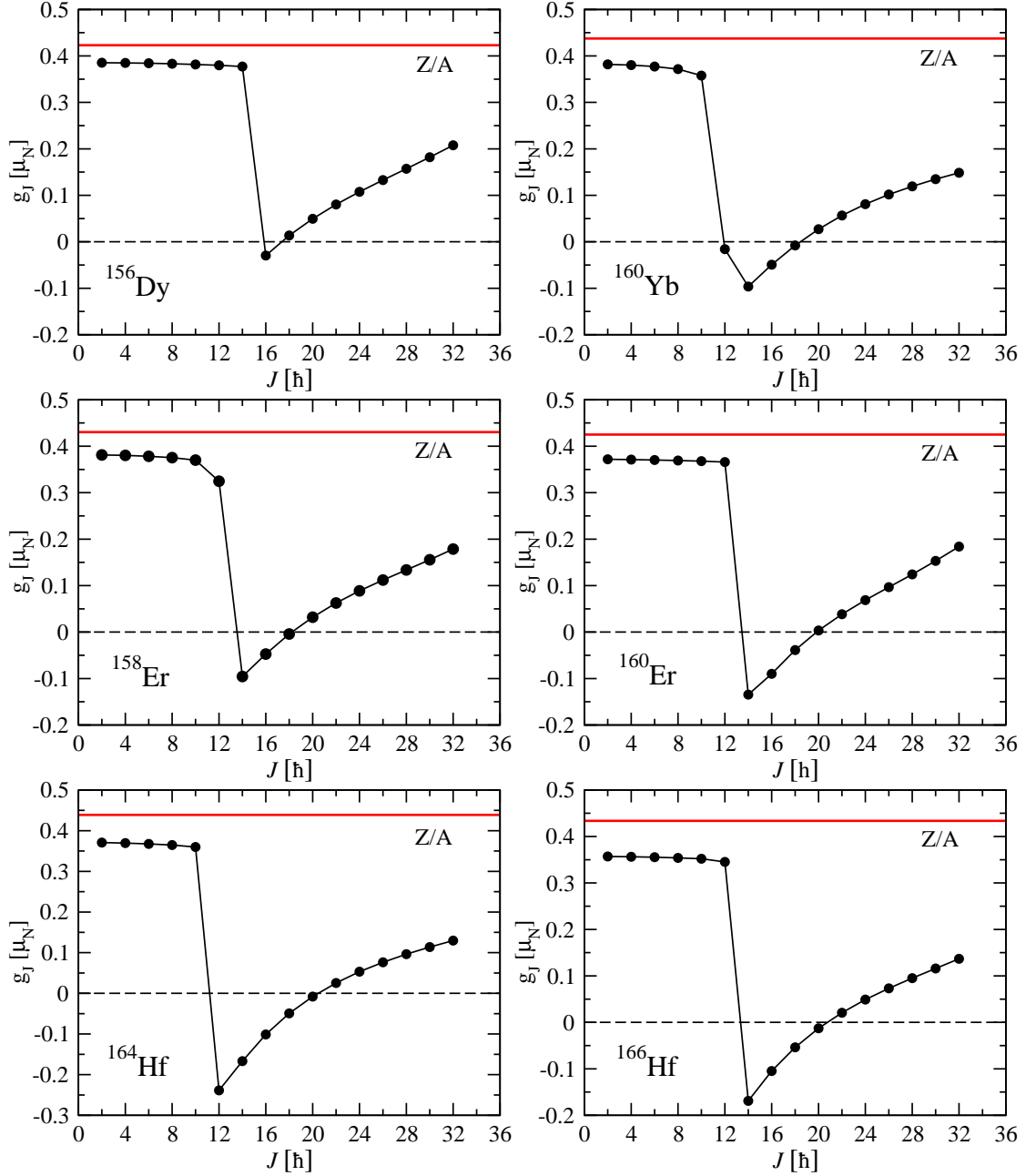}
\end{center}
\caption{(Color online) The gyromagnetic factor given in units of nuclear magneton, calculated for yrast states, is represented as function of angular momentum. }  
\label{Fig.8}
\end{figure}
Equations (5.7) and (5.8) can be used for calculating the gyromagnetic factor for an yrast state of angular momentum $J$. Indeed, from the expression of the magnetic moment
\begin{equation}
\vec{\mu}=g_c\vec{J}_c+g_f\vec{J}_f\equiv g_J\vec{J},
\end{equation}
one easily derive the following expression for $g_J$:
\begin{equation}
g_J=g_c+\frac{g_f-g_c}{2}\left[1+\frac{\tilde{J}_f(\tilde{J}_f+1)-\tilde{J}_c(\tilde{J}_c+1)}
{J(J+1)}\right],
\end{equation}
where $g_c$ and $g_f$  denote the gyromagnetic factor of the core and fermionic system.
For the core we consider
\begin{equation}
g_c\approx \frac{Z_c}{A_c},
\end{equation}
with $Z_c$ and $A_c$ denoting the charge and atomic number characterizing the core. Taking into account that the intruder  state is $i_{13/2}$ and this is occupied by neutrons, the corresponding gyromagnetic factor is:
\begin{equation}
g_f=\frac{g^n_s}{13}\approx -0.2943 \mu_N,
\end{equation}
with $\mu_N$ being the nuclear magneton and $g^n_s$ the gyromagnetic factor for the neutron spin. 
The gyromagnetic factor plotted in Fig.9 as a function of $J$, reflects the nature of the yrast states. Indeed, before the intersection of the $0qp$ and $2qp$ bands, $g_J$ is independent of $J$
and  very close to the rotational value $Z/A$. When the intersection of the two bands takes place, $g_J$ has a big jump to a negative value, which confirms the $2qp$ character of the band which follows. This feature persists only for a few states and then the core contribution starts to be dominant which results in having a positive value for $g_J$. The curve allure suggests a quadratic $J$ dependence for $g_J$. Except for $^{160}$Yb, the curve has two branches, one constant corresponding to the $g$ band and one quadratically increasing with $J$, which corresponds to the 
$S$ band. In the case of $^{160}$Yb, after the crossing point, the curve decreases a little and then starts increasing. Note that because the magnetic moment carried by the core and fermions have opposite orientations and moreover the core contribution is an increasing function of $J$, there is a critical value of $J$ where the total magnetic moment and therefore the gyromagnetic factor are vanishing. For the maximum angular momentum considered here, the gyromagnetic factor reaches half of the rotational value, i.e. about 0.2 $\mu_N$. 

Before closing this section we would like to spend a few lines on comparing our approach with two
other formalisms.
At a superficial glance one may think that the formalism presented here is similar to that of 
Refs.\cite{Ikeda,Ikeda1}. Therefore a fair comparison of the two procedures is necessary. First we mention that the mean-field determining the single-particle basis is deformed while that used in Refs.\cite{Ikeda,Ikeda1} is spherical. Therefore, an angular momentum projection operation for the deformed BCS states is necessary in our case. In the quoted reference, the coherent state is defined in terms of deformed quadrupole bosons which makes difficult the evaluation of the overlap matrix elements. Indeed, the matrix elements between rotated intrinsic states are approximated and then parametrized.
By contrast, the coherent state used here corresponds to spherical quadrupole bosons and moreover the matrix elements are analytically calculated. Here the pairing interaction is effective for nucleons moving outside the phenomenological core, while in Ref. \cite{Ikeda,Ikeda1} only the pairs from the core and those from outside the core interact with each other. Moreover, the pairing interaction is treated in the particle representation which in fact leads to a parametrization of the corresponding matrix elements. Consequently, the Hamiltonian is diagonalized in a basis with a fixed number of particles. Since a BCS formalism for the deformed single-particle states is used, in the present case, the number of particles is conserved only in average. The numbers of the parameters employed by the two formalisms are also different: seven for Ref.\cite{Ikeda,Ikeda1} and five in the present approach.
One may conclude that although the two approaches have some common features, they are  essentially different.
A similar single-particle basis is used in Ref.\cite{Hama} but in a different context. Indeed, 
using a particle-rotor formalism the dependence of the interaction of the lowest two bands on the 
degree of filling the shell is studied pointing out an oscillating behavior.

\renewcommand{\theequation}{6.\arabic{equation}}
\section{Conclusions}
\label{sec: level6}

In the previous sections a  semiphenomenological formalism for the description of the backbending phenomenon was proposed.   A  model Hamiltonian associated with  a set of   interacting particles moving in a deformed mean-field coupled to a phenomenological core described in terms of quadrupole boson operators  is treated in a product space of angular momentum projected states.  
The pairing interaction of neutrons moving in a deformed mean-field is treated by the BCS formalism. The model states for the ground-state band are obtained by angular momentum projection of the deformed product state $|BCS\rangle_d\psi_c$,
 while  the $S$ band states are projected out from the intrinsic $K=1$ two quasiparticle states $J_+\alpha^{\dagger}_{jk}\alpha^{\dagger}_{j-k}|BCS\rangle_d\psi_c$.  
The substate $|jk\rangle$ is chosen such that the corresponding quasiparticle energy is minimum.
Projected states of $g$ and $S$ bands are not mutually orthogonal. Diagonalizing the overlap matrix, one defines an orthogonal basis for treating the model Hamiltonian.
The parameters involved were fixed by a fitting procedure described in the previous section. The lowest  Hamiltonian eigenvalues in the orthogonal basis defines the yrast band. The first energy levels
originate from the projected states of the $0qp$ state mentioned above while, starting from a critical angular momentum they are mainly of a $2qp$ nature.

The experimental backbending shape of the moment of inertia versus angular frequency squared is fairly well reproduced by our results. The data considered in our calculations refer to energy levels with angular momentum up to 26-28. A measure of the agreement quality is the $r.m.s.$ value characterizing the deviations of the calculated energies from the corresponding experimental data. This quantity is about 30keV or less.

A detailed analysis of the effect coming from the $H_{J_fJ_c}$ term concerning the neutron pair breaking as well as the alignment of individual angular momentum to the core angular momentum, is presented. The pair breaking takes place for $J=10,12$ and the maximum alignment is settled for $J$
larger than 20. Since the de-paired neutrons carry a projection $K=1$, a full alignment is not possible. As shown in Fig.5 for $J\geq 20$ the angular momentum defect $\Delta J$ reaches a
 plateau with $\Delta J=2$. Also observed is an abrupt change of the gyromagnetic factor from positive to negative values in the band crossing region, which is consistent with the change in structure of the yrast band, going from rotational to $2qp$ character.

We may conclude that the present formalism is able to account quantitatively for the main features of the backbending phenomenon in the rare earth nuclei.

Although effects like pair breaking, bands crossing and angular momentum alignment have been explained, within some cranking formalisms, by several authors, the present approach provides a consistent description for the mentioned effects pointing out several interesting features which will be enumerated below.

\begin{itemize}

\item{Although we use a spherical projected particle-core basis the two components, particles and
core, are deformed. Working with states of good angular momenta is an advantage over the cranking methods where the states in the crossing region exhibit a large dispersion for angular momentum}.
\item{ The effects of the $qQ$ and $\vec{J}_f\cdot\vec{J}_c$ terms on the backbending are discussed and specific contributions are identified.} 
\item{ It is shown that the spin-spin interaction simulates in the laboratory frame the Coriolis force which is active in the intrinsic frame.}
\item{By the bands crossing the core angular momentum, $\tilde{J}_c$, varies only by a few units, $2-4 \hbar$. The minimal variation is of $2\hbar$ and is recorded for $^{160}$Yb. Due to this feature one expects a nonvanishing $E2$ transition between the states lying in the vicinity of the crossing point.}
\item{We depicted one case, $^{160}$Yb, where after the bands crossing point, the angular momenta carried by fermions and the core are close to each other.}
\item{(6.)The curves associated with $\tilde{J}_c$ and $\tilde{J}_f$, in Fig.5, cross each other at a total angular momentum equal to 22 for $^{156}$Dy, $^{158}$Er, 24 for $^{160}$Yb, $^{160}$Er and
26 for $^{164,166}$Hf.}
\item{The maximum alignment is reached in the region of $J_c>J_f$, where the defect $\Delta J$ is equal to $2\hbar-3\hbar$. This is a reflection of the $K=1$ nature for the $S$ band.}
\end{itemize}

Due to the above-mentioned aspects one may say that although the present paper addresses a relatively old subject the  proposed formalism unveils alternative features of the backbending phenomenon.
 
Before closing, we present a few perspectives of the present approach. Indeed, the results encourage us to extend the restricted model space by adding to the particle factor the proton state
 $h_{11/2}$ which is suspected to be responsible for the second backbending. Another possible extension refers to the collective factor states, by adding to the coherent state considered here
the model states for beta and gamma bands used  by the coherent state model.
 It is well known that the energy spectra of these bands comprise more irregularities than  the ground state  band. It is an open question whether such anomalies could be also interpreted as the interaction with other bands of a different nature. In this way we could describe a multibackbending phenomenon showing up in the yrast and non-yrast bands.

{\bf Acknowledgment.} This work was supported by the Romanian Ministry for Education Research Youth and Sport through  CNCSIS Project No. ID-1038/2008.

\renewcommand{\theequation}{A.\arabic{equation}}
\section{Appendix A}
\label{sec: level7}
The analytical expressions for the terms $A_1(M,J)$ and $\tilde{B}_n(M,J)$ involved in Eq.(2.10)
are as follows:
\begin{eqnarray}
\tilde{B}_{n}(M,J)&=&\delta_{J,0}+\sum_{m=2, \atop m=even}^{M}I_{b}(J,m)\cos{\left[\frac{\pi}{M+1}m\cdot n\right]},\\
I_{b}(J,m)&=&\int_{0}^{\pi}d\beta\sin{\beta}P_{J}(\cos{\beta})e^{-im\beta},\\
A_{1}(M,J)&=&\frac{\prod_{k=1}^{J}(M-J+2k)}{\prod_{k=0}^{J}(M-J+2k+1)}.
\end{eqnarray}
The integral $I_{b}(J,m)$ can be analytically determined, and the result for the present case $J=$even and $m=$even is:
\begin{equation}
I_{b}(J,m)=\left\{\begin{array}{l}\displaystyle{-2\frac{\prod_{k=0}^{\frac{J}{2}-1}\left[m^{2}-(2k)^{2}\right]}{\prod_{k=0}^{\frac{J}{2}}\left[m^{2}-(2k+1)^{2}\right]},\,\,\,m\geq J},\\
0,\,\,\,m<J.\end{array}\right.
\end{equation}

\renewcommand{\theequation}{B.\arabic{equation}}
\section{Appendix B}
\label{sec: level8}
The state obtained by applying a pair of time reversed quasiparticle operators on a BCS function can be written as a linear combination of states with a definite angular momentum $J$, with $J$ running from 0 to 24, and a projection on $z$ axis $K=0$:
\begin{equation}
\alpha_{jk}^{\dagger}\alpha_{j-k}^{\dagger}|BCS\rangle_{d}=\sum_{J}C_{J}^{jk}|J,0\rangle.
\end{equation}
Acting with an angular momentum projection operator on this state is equivalent to selection of only one $J$ component from the above linear combination, and rotating the projection to the value $M$:
\begin{equation}
P_{M_{f}0}^{J_{f}}\alpha_{jk}^{\dagger}\alpha_{j-k}^{\dagger}|BCS\rangle_{d}=C_{J_{f}}^{jk}|J_{f},M_{f}\rangle.
\end{equation}
The angular momentum states $|J,M\rangle$ are orthogonal and normalized to unity. Then the norm of this projected function is the reciprocal of the amplitude $C_{J_{f}}^{jk}$: 
\begin{equation}
\left(C_{J_{f}}^{jk}\right)^{2}=~_{d}\langle BCS|\alpha_{-k}\alpha_{k}P_{00}^{J_{f}}\alpha_{k}^{\dagger}\alpha_{-k}^{\dagger}|BCS\rangle_{d}.
\end{equation}

Now if we apply on the state (B.1) a raising angular momentum operator,
\begin{equation}
J_{+}\alpha_{jk}^{\dagger}\alpha_{j-k}^{\dagger}|BCS\rangle_{d}=\sum_{J}C_{J}^{jk}\sqrt{J(J+1)}|J,1\rangle,
\end{equation}
and then project an angular momentum from  the resulting function, one obtains exactly the $K=1$ $2qp$ projected fermionic function:
\begin{equation}
P_{M_{f}1}^{J_{f}}J_{+}\alpha_{jk}^{\dagger}\alpha_{j-k}^{\dagger}|BCS\rangle_{d}=C_{J_{f}}^{jk}\sqrt{J_{f}(J_{f}+1)}|J_{f},M_{f}\rangle,
\end{equation}
whose normalization factor is readily obtained:
\begin{equation}
\left(\mathcal{N}_{J_{f};1}^{jk}\right)^{-2}=J_{f}(J_{f}+1)~_{d}\langle BCS|\alpha_{-k}\alpha_{k}P_{00}^{J_{f}}\alpha_{k}^{\dagger}\alpha_{-k}^{\dagger}|BCS\rangle_{d}.
\end{equation}

Making use of Eqs. (B.5) and (B.6), one can also derive the expressions for other  m.e. which we need in our calculations. For example the overlap of the $0qp$ and $2qp$ projected states  can be expressed as
\begin{equation}
_{d}\langle BCS|P_{01}^{J_{f}}J_{+}\alpha_{jk}^{\dagger}\alpha_{j-k}^{\dagger}|BCS\rangle_{d}=\left[N_{J_{f}}^{BCS}\mathcal{N}_{J_{f}1}^{jk}\right]^{-1}.
\end{equation}

In order to calculate the quantities  $_{d}\langle BCS|\alpha_{j-k}\alpha_{jk}P_{00}^{J_{f}}\alpha_{jk'}^{\dagger}\alpha_{j-k'}^{\dagger}|BCS\rangle_{d}$ first are calculated the corresponding matrix elements with particle operators $c^{\dagger}_{jk}(c_{jk})$ instead of quasiparticle ones $\alpha_{jk}^{\dagger}(\alpha_{jk})$, for which we use the same method as in the case of the norm of the projected BCS state. Having these determined, the desired matrix elements are easily obtained  by applying a canonical transformation to the particle operators:
\begin{gather}
_{d}\langle BCS|\alpha_{j-k}\alpha_{jk}P_{00}^{J_{f}}\alpha_{jk'}^{\dagger}\alpha_{j-k'}^{\dagger}|BCS\rangle_{d}=\frac{1}{U_{jk}^{2}U_{jk'}^{2}}~_{d}\langle BCS|c_{j-k}c_{jk}P_{00}^{J_{f}}c_{jk'}^{\dagger}c_{j-k'}^{\dagger}|BCS\rangle_{d}\nonumber\\
-\frac{V_{jk'}V_{jk}}{U_{jk'}U_{jk}}(-)^{k+k'}~_{d}\langle BCS|P_{00}^{J_{f}}|BCS\rangle_{d}-\frac{V_{jk}}{U_{jk}U_{jk'}^{2}}(-)^{j-k}~_{d}\langle BCS|P_{00}^{J_{f}}c_{jk'}^{\dagger}c_{j-k'}^{\dagger}|BCS\rangle_{d}\nonumber\\
-\frac{V_{jk'}}{U_{jk'}U_{jk}^{2}}(-)^{j-k'}~_{d}\langle BCS|c_{j-k}c_{jk}P_{00}^{J_{f}}|BCS\rangle_{d},
\end{gather}
where $U$ and $V$ are the occupation parameters from the BCS equations.

\end{document}